\documentclass[preprint,journal]{vgtc}            % preprint (journal style)

\usepackage{balance}   % Allows balancing the columns on the last page
\usepackage{xcolor}
\usepackage{colortbl}
\usepackage{amsmath}
\usepackage{longtable}
\usepackage{float} 
\usepackage{bm}
\onlineid{0}

\vgtccategory{Research}

\vgtcpapertype{please specify}

\title{Affective Color Scales for Colormap Data Visualizations}

\author{%
  Halle C. Braun,
  \authororcid{Kushin Mukherjee}{0000-0001-5013-6983},
  \authororcid{Seth R.\ Gorelik}{0000-0002-9186-1767}, and
  \authororcid{Karen B.\ Schloss}{0000-0003-4833-4117}
}

\authorfooter{
  \item
  	Halle C. Braun is with the University of Wisconsin--Madison.
  	E-mail: hallebraun43@gmail.com
  \item
  	Kushin Mukherjee is with Stanford University.
  	E-mail: kushinm@stanford.edu

  \item Seth R. Gorelik is with Woodwell Climate Research Center. 
  	E-mail: sgorelik@woodwellclimate.org

      \item Karen B. Schloss is with the Univeristy of Wisconsin--Madison.
  	E-mail: kschloss@wisc.edu    
}

\abstract{
 Research on affective visualization design has shown that color is an especially powerful feature for influencing the emotional connotation of visualizations. Associations between colors and emotions are largely driven by lightness (e.g., lighter colors are associated with positive emotions, whereas darker colors are associated with negative emotions). Designing visualizations to have all light or all dark colors to convey particular emotions may work well for visualizations in which colors represent categories and spatial channels encode data values. However, this approach poses a problem for visualizations that use color to represent spatial patterns in data (e.g., colormap data visualizations) because lightness contrast is needed to reveal fine details in spatial structure. In this study, we found it is possible to design colormaps that have strong lightness contrast to support spatial vision while communicating clear affective connotation. We also found that affective connotation depended not only on the color scales used to construct the colormaps, but also the frequency with which colors appeared in the map, as determined by the underlying dataset (data-dependence hypothesis). These results emphasize the importance of data-aware design, which accounts for not only the design features that encode data (e.g., colors, shapes, textures), but also how those design features are instantiated in a visualization, given the properties of the data. 
}


\teaser{
  \centering 
  \includegraphics[width=.85\linewidth, alt={}]{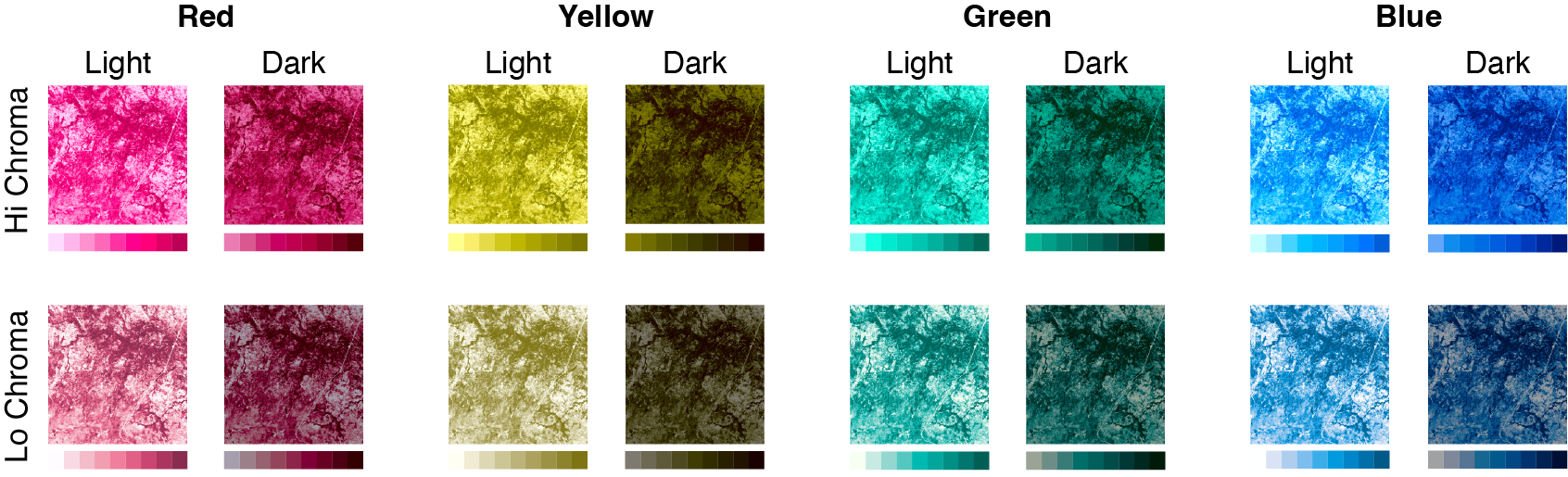}
  \caption{Example colormaps from Exp. 1, which represent raster data of aboveground live woody biomass \cite{harris2021}. The color scales shown below each colormap were generated in Color Crafter \cite{smart2020} by adjusting hue (red, yellow, green, and blue), lightness (light or dark), and chroma (high chroma or low chroma) within the tool. Here, all 16 color scales were applied to the same underlying dataset, but in the experiment, participants saw each color scale applied to a unique dataset (balanced across participants).}
  \label{fig:teaser}
}

\graphicspath{{figs/}{figures/}{pictures/}{images/}{./}} % where to search for the images

\usepackage{tabu}                      % only used for the table example
\usepackage{booktabs}                  % only used for the table example
\usepackage{lipsum}                    % used to generate placeholder text
\usepackage{mwe}                       % used to generate placeholder figures

\usepackage{mathptmx}                  % use matching math font

\begin{document}

%%%%%%%%%%%%%%%%%%%%%%%%%%%%%%%%%%%%%%%%%%%%%%%%%%%%%%%%%%%%%%%%
%%%%%%%%%%%%%%%%%%%%%% START OF THE PAPER %%%%%%%%%%%%%%%%%%%%%%
%%%%%%%%%%%%%%%%%%%%%%%%%%%%%%%%%%%%%%%%%%%%%%%%%%%%%%%%%%%%%%%%

\firstsection{Introduction}

\maketitle
In colormap data visualizations, gradations of magnitude are mapped onto gradations of color to illustrate spatial patterns within a dataset. A large body of research on colormap (a.k.a. heatmap) design has focused on which kinds of color scales (a.k.a. color ramps) are most effective for revealing patterns in data \cite{antes1990, bujack2018, moreland2009, nardini2021,smart2020, brewer1994, reda2018, reda2020, reda2021, rogowitz1996, rogowitz1998, zhou2016, zeng2021} and how to assign the endpoints of color scales (e.g., light vs. dark) to endpoints of data magnitude (e.g., low vs. high) to help observers interpret the meaning of colors in colormaps \cite{cuff1973, mcgranaghan1989, schloss2019, sibrel2020, schoenlein2023, soto2023}. The aforementioned work focused on people's ability to detect and analyze patterns of data, stemming from the general emphasis in visualization research on accuracy, conciseness, and clarity (c.f. \cite{lan2024}). 
However, another aspect of visualization design has received a recent surge of attention---affective, or emotional, aspects of design \cite{bartram2017, anderson2022, blair2025, samsel2018ArtAffect, zeller2022, schombs2024, lan2021Kineticharts, lan2021, lan2022, zhang2021}---which can influence engagement, recall, and aesthetic response (see Lan et al.\cite{lan2024} for a review). In the domain of color, previous work has focused on understanding affect of color \textit{palettes} (i.e., discrete colors to visualize categorical data) \cite{bartram2017, anderson2022, blair2025}, and explored using palettes from paintings to imbue colormaps with affective connotations \cite{samsel2018ArtAffect}. Here, we aimed to understand factors that influence affective connotation of colormaps for visualizing continuous data. 

The discussion of affect in visualization is often couched in terms of whether a design is embellished to be `emotional' or kept minimalistic to be `neutral' (c.f. \cite{lan2024}), but an alternative lens is that all designs can be situated in a continuous space of affective connotation. After all, seemingly neutral `minimalistic' grayscale visualizations are reportedly linked to negative emotions \cite{blair2025, zhang2021}.
From this perspective, research on affective visualization design aims to understand factors that link emotion and design.
These links can take two possible forms: \textit{affective connotations}---associations observers have between the design and emotional concepts or \textit{affective state}---emotions observers feel when they look at visualizations. Evidence suggests that color is among the strongest features linking emotion to visualization, compared to other features such as data patterns (e.g., positive/negative slope) or chart type (e.g., scatterplot, bar chart) \cite{blair2025}.  

Previous work on links between color and emotion has largely focused on individual colors \cite{collier1996, dandrade1974, wexner1954, kaya2004, valdez1994, suk2010, wilms2018} (see Jonauskaite and Mohr \cite{jonauskaite2025} for a review), color pairs \cite{ou2004_pairs, ou2012}, faces/scenes  \cite{thorstenson2018, pazda2024}, or color palettes for visualizations of categorical data \cite{bartram2017, anderson2022, blair2025}, but less so on visualizations of continuous data \cite{samsel2018ArtAffect, zhang2021}. A general theme across prior studies is that color-emotion links are largely driven by lightness and chroma. For example, \textit{positive} colors are light and high in chroma, \textit{negative} colors are dark and low in chroma, and \textit{calm} or \textit{less angry} colors are light and low in chroma \cite{bartram2017, schloss2020, dandrade1974, wright1962, dael2016, jonauskaite2019mood}. Although hue plays a role to some extent (and for some emotions more than others), many hues have similar affective connotation, especially if the colors are light \cite{bartram2017, schloss2020}.

\begin{figure}[tb]
 \centering
 \includegraphics[width=1.0\columnwidth]{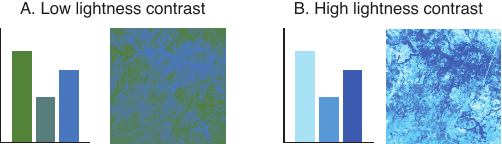}
 \caption{Visualizations of categorical data (bar chart) and continuous data
(colormap) using colors that have (A) low or (B) high lightness contrast.}
 \label{fig:contrast}
 \vspace{-6mm}
\end{figure}

The strong link between lightness and emotion has interesting implications for visualization design. For visualizations of categorical data (e.g., bar charts), in which data magnitude is encoded using spatial channels (e.g., position, length) and categories are encoded using color, colors of similar lightness may signal strong affective connotation without harming the observer's ability to find key patterns in the data. For example, comparing the heights of the bars in Fig. \ref{fig:contrast}A (low lightness contrast) seems as easy as in Fig. \ref{fig:contrast}B (high lightness contrast). But, for visualizations of continuous data, in which data magnitude is encoded using dimensions of color, minimizing lightness variability can pose problems because lightness contrast is important for spatial vision. For example, observing spatial structure in the low lightness contrast colormap in Fig. \ref{fig:contrast}A is harder than in the high lightness contrast colormap Fig. \ref{fig:contrast}B. This difficulty is because the visual system relies on variation in lightness to detect spatial structure in data \cite{ware1988}, particularly for high spatial frequencies \cite{rogowitz1996} (e.g., fine details in a geographical map). Generally, spatial vision (e.g., contrast sensitivity in high spatial frequencies, orientation sensitivity, pattern and shape recognition) is worse when colors are isoluminant \cite{devalois1990, devalois1993, rogowitz1996, kindlmann2002, livingstone1987}. Yet, introducing lightness variability to support spatial vision will also introduce variability in affective connotation because of the strong link between lightness and emotion. Thus, we asked, \textbf{is it possible to produce colormap data visualizations with strong lightness contrast that also have systematic affective connotations?}

We addressed this question by investigating color-emotion associations for colormaps that differ systematically in lightness, chroma, and hue while preserving lightness contrast (Fig. \ref{fig:teaser}). We created the colormaps using color scales generated by Color Crafter \cite{smart2020}, a tool that allows for flexible and automatic production of designer-grade color scales (such as ColorBrewer color scales \cite{harrower2003}). We then applied the color scales to raster data of aboveground live woody biomass \cite{harris2021} to create colormap data visualizations (Fig. \ref{fig:teaser}). We used a real-world dataset as opposed to synthetic data to support ecological validity of our experiment while avoiding landmarks that would be recognizable in the geographical regions (henceforth, we use `subtile' to refer to these geographical regions). Based on results from a pilot study, we also ensured that the subtiles had similar affective connotation when presented in grayscale so observed effects could be attributed to color and not to spatial aspects of the data. In Exp. 1, the values in the underlying dataset were well-distributed so that each color of a color scale was well-represented in the corresponding colormaps. In Exp. 2, we shifted the underlying data distributions to test if affective connotation depended not only on the colors in the color scale, but also on the relative frequency with which each color appeared in the colormap (data-dependence hypothesis).

\textbf{Contributions.} In this paper we make three main contributions: (1) We provide evidence that colormaps with large lightness variation can have systematic affective connotations. (2) We provide support for the data-dependence hypothesis, which emphasizes the need for data-aware design---affective connotation depends not only on the color scale used to construct a colormap, but also on the relative proportion of colors determined by the underlying dataset. (3) We describe affective connotations of colormaps generated from 16 different color scales and how connotation is modulated by the underlying dataset, which can inform the use of color in future research and visualization design. In the General Discussion, we consider open questions to build on this foundation towards a comprehensive understanding of affective visualization design for colormaps.

\section{Background}

\subsection{Affective visualization design} 
As defined by Lan et al. \cite{lan2024}, \textit{affective visualization design} concerns data visualization designed to communicate and influence emotion. It is part of a larger class of \textit{affective design}, which also includes designs to visualize aspects of emotion (e.g., measures of heart rate) and studies of how priming an observer's affective state influences their evaluations and interpretations of visualizations. Lan et al. \cite{lan2024} reviewed several kinds of methods in affective visualization design, including (1) `sensation', pertaining to links between perceptual experiences and emotion, (2) `narrative', pertaining to storytelling techniques used to communicate data, (3) `behavior', pertaining to manipulating/operating on data, creating/building data visualizations, walking/wandering through visualizations, and (4) `context', pertaining to the surrounding environment when people view or interact with data. `Sensation' is the most pertinent factor for the present study, so the rest of this section will focus on links between perceptual features and emotions. 

When considering links between perceptual features and emotions, there is a qualitative distinction between \textit{affective connotation} (i.e., conceptual association) between perceptual features and emotions, and \textit{affective state} (i.e., feelings) elicited from perceptual features. For example, people strongly associate high chroma reds with the concept of anger, but that does not necessarily mean they \textit{feel} anger when they see high chroma reds \cite{schloss2020, jonauskaite2025}. This distinction is perhaps best illustrated in the domain of music. Individuals with \textit{musical anhedonia} do not experience pleasure while listening to music \cite{kathios2024, loui2017, belfi2020}, but their ability to identify emotional associations with music is similar to typical listeners \cite{satoh2011} (see \cite{kathios2024, loui2017, belfi2020} for potential mechanisms).  

Different kinds of methods are used for assessing affective connotation and affective state. Methods for assessing affective connotation are relatively straightforward and include having observers rate associations between perceptual features and emotion-related concepts, match visualizations to emotion-related concepts, or assess which features observers choose when designing visualizations to communicate particular emotion concepts. In contrast, methods for assessing affective state in response to perceptual stimuli are more complicated and researchers have concerns about interpreting the measures\cite{barrett2021, jonauskaitePICS}. One method is having participants self-report on their emotions through ratings or verbal descriptions, but there are concerns that observers may not distinguish between felt emotions and associations with emotions when self-reporting on affective state and that ratings may be influenced by the wording or images used in emotion scales. A second class of methods involve physiological measures, but physiological measures can contradict self-report measures and it is unclear how to interpret such conflicts. Given complications surrounding the assessment of felt emotions in response to visual features, this study focuses on affective connotation of colormap visualizations.

\subsection{Correspondences between color and emotion}
Researchers have been studying links between individual colors and emotion for over 100 years (see Jonauskaite and Mohr \cite{jonauskaite2025} for a review). Evidence suggests that associations between single colors and emotion are largely driven by lightness and chroma, with some systematic effects of hue \cite{wright1962, dandrade1974, schloss2020, jonauskaite2019mood, dael2016}. For example, Schloss et al. \cite{schloss2020} asked participants to rate emotional associations for single colors that varied independently in hue angle, lightness, and chroma (similar to our approach in the present study). They found that lighter, higher chroma colors were happier (and less sad) than darker, lower chroma colors. They assessed effects of hue by testing for differences across the red/green and yellow/blue dimension (a* and b* in CIELAB space, respectively) and found that bluer colors were happier than yellower colors, particularly for darker colors. This finding may seem surprising given the conventional notion that yellows are happier than blues, but that notion is likely an artifact of the most typical yellows being lighter than the most typical blues (and colors of yellow hue crossing the category boundary from yellow to brown as they darken and are less likely to be called yellow) \cite{schloss2020, dandrade1974, jonauskaite2025}. The dominance of lightness in color-emotion associations may explain why red-green dichromatic observers have similar color-emotion associations as observers with typical color vision \cite{jonauskaite2021dichrom}---both groups can perceive lightness differences.

In their study of affective connotation of color palettes, Bartram et al. \cite{bartram2017} also found strong effects of lightness and chroma. They began by generating candidate colors for a design study by analyzing the color palettes of images tagged with eight emotion concepts that span the pleasure (valence) and arousal (activation) axes of the pleasure, arousal, dominance (PAD) model of affect \cite{russell1980,posner2005}:  \textit{calm}, \textit{exciting}, \textit{serious}, \textit{playful}, \textit{positive}, \textit{negative}, \textit{disturbing}, and \textit{trustworthy}.  
Then, they had users compose palettes for visualizations that best communicated each emotion. Colors selected for the terms \textit{positive}, \textit{playful}, and \textit{calm} were lighter than the colors selected for \textit{disturbing}, \textit{serious}, and \textit{negative}. In terms of chroma, colors selected for \textit{exciting}, \textit{playful}, and \textit{positive} were high in chroma, and colors for \textit{calm} were low in chroma. There were also some effects of hue, such as \textit{calm} being linked to cooler hues (blues and greens) and \textit{exciting} being linked to warmer hues (red, oranges, yellows). Bartram et al. \cite{bartram2017} found similar results when they asked observers to rank visualizations for each affective term. 

Extending from Bartram et al. \cite{bartram2017}, Anderson and Robinson \cite{anderson2022} investigated the effects of affective congruence between color palette and data topic on interpretations and evaluations of choropleth maps. They chose color palettes and topics that were 
\textit{positive}, \textit{negative}, \textit{calm}, or \textit{exciting} (palettes from\cite{bartram2017}) and paired them so they were congruent (e.g., high chroma colors with positive topics like `kinds of desserts'; low chroma colors with negative topics like `methods of homicide') or incongruent (i.e., opposite color/topic pairing). Congruency was defined with respect to the palette as a whole, without considering semantic discriminability of the colors \cite{schloss2021, mukherjee2022} or optimal color-concept assignments \cite{lin2013, schloss2018, schloss2021, mukherjee2022}. As such, they did not find effects of affective congruence on interpretive tasks that required mapping concepts to individual colors. Yet, they did find that affectively congruent visualizations had amplified emotional connotation (e.g., maps of pleasurable topics were judged as especially pleasurable when the colors were positive rather than negative) and were rated as being more appropriate. These results suggest there are important downstream implications of the affective connotation of color in visualizations.  

Most recently, Blair et al. \cite{blair2025} assessed self-reported affective state (felt emotion) in response to color, as well as other aspects of visualization design: chart type, data trend, data variance, and data density. They found that all of these factors contributed to affective state, but color was an especially powerful cue. Moreover, reports of affective state were similar for abstract color palettes (i.e., colored squares) and colors presented in scatterplot visualizations. These findings suggest that affective response to color may be agnostic to presentation format and design may not need to be data-aware, at least for categorical visualizations where all palette colors can have similar affective connotation. We will return to this issue in Exp. 2 and the General Discussion.

Research on links between color and emotion for visualization primarily focused on categorical data, but there are some exceptions. Zhang et al. \cite{zhang2021} studied links between emotional response to colormaps and visualization comprehension and found that colormaps eliciting more positive affective responses supported greater comprehension. They did not report on what aspects of color appearance influenced affective response, but their results imply that colormaps containing lightness contrast may evoke specific emotions.

To design affectively connotative colormaps, Samsel et al. \cite{samsel2018ArtAffect} took inspiration from colors in paintings. Their process involved choosing a painting that had a desired affective connotation, selecting representative colors from the painting to form a palette, using the palette as input into \href{https://sciviscolor.org/colormovesapp}{ColorMoves} to create a color scale, and then applying the color scale to the dataset to produce the colormap visualization. Although they did not assess affective connotation in the resulting colormaps, the premise was that the colormaps would share the connotation of paintings from which the colors were sourced. 

Building on this prior work, we aimed to understand the factors influencing affective connotation of colormap data visualizations, focusing on the dimensions of color and properties of the underlying dataset.

\section{Experiment 1}
Exp. 1 assessed color-emotion associations for colormaps that differed systematically in hue, lightness, and chroma. The underlying datasets were selected such that all colors in the color scales were well-represented in the colormap visualizations. 

We hypothesized that although all the colormaps varied in lightness, their overall lightness level would drive affective connotation. Moreover, based on the findings for single colors \cite{schloss2020, dandrade1974, wright1962, jonauskaite2025}, we hypothesized that positive affective terms would be associated with lighter, higher chroma, and bluer colormaps, whereas negative affective terms would be associated with darker, lower chroma, and yellower colormaps. We also hypothesized that the concept \textit{calm} would be associated with cooler, lighter, and lower chroma colormaps \cite{bartram2017}. The experimental stimuli, data, and analysis scripts for both experiments in this study can be found in our GitHub repository\footnote{\url{https://github.com/SchlossVRL/color_scales_affect}}. 

\begin{figure}[tb]
 \centering
 \includegraphics[width=0.95\columnwidth]{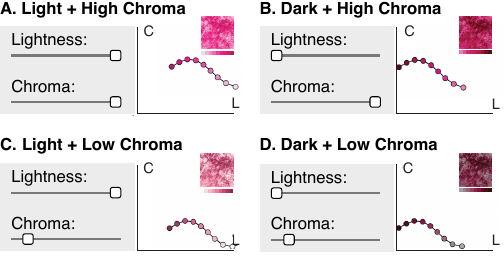}
 \caption{Illustration of how color scales were generated in Color Crafter by adjusting lightness and chroma sliders (note, in Color Crafter the scale controlling chroma is labeled `saturation').  The charts rightward of slider pairs show Color Crafter's charts of the 9-point color scale plotted in terms of lightness (L) and chroma (C) of CIELCh space. The inset in each plot shows an example colormap produced from that color scale.}
 \label{fig:colorcrafter}
 \vspace{-6mm}
\end{figure}

\subsection{Methods}
\subsubsection{Participants}
The participants ($n=35$) were undergraduates from the University of Wisconsin--Madison (UW--Madison) who received extra credit in their Psychology course (mean age: 18.8, range: 18-23). They provided demographic information in free-response text boxes. Their genders included 16 women and 19 men (no other genders reported). Their race/ethnicity included 27 White, 5 Asian, 2 Hispanic, 1 Black, 1 Arab, 1 African-American, and 1 Middle Eastern (the number of entries exceeds the number of participants because some reported $>1$ race/ethnicity). For all experiments in this study, all participants gave informed consent and the UW--Madison IRB approved the protocol.

Our target sample was $n=32$, established as follows. Based on the sample size of $n=20$ in Exp. 2 of \cite{schloss2020} (analogous to the present experiment but for single colors), we sought to collect data from at least 20 participants. To reach $n \geq 20$ and to accommodate the 16 conditions in the Latin square design that balanced the color scale and dataset assignments across participants (Section \ref{sec:exp1_design_displays}), we doubled the Latin square resulting in target $n=32$. We ran $n=35$ to accommodate excluding 3 participants with atypical color vision determined by self-report questions or errors on more than 2 of 11 digital Ishihara plates presented at the end of the study (criteria determined in advance).

\subsubsection{Design and displays} \label{sec:exp1_design_displays}

 Based on the approach in Schloss et al. \cite{schloss2020} for single colors, we constructed colormaps generated from 16 color scales, comprising the orthogonal combination of 4 hues (red, yellow, green, and blue) $\times$ 2 lightness levels (light, dark) $\times$  2 chroma levels (low, high) (Fig.\ref{fig:teaser}). We generated the color scales in Color Crafter \cite{smart2020}, detailed below. Each participant rated the association between each of 16 colormaps and each of 8 emotion concepts from Bartram et al. \cite{bartram2017} (\textit{positive}, \textit{playful}, \textit{exciting}, \textit{negative}, \textit{serious}, \textit{disturbing}, \textit{calm}, and \textit{trustworthy}) twice, resulting in 256 trials.

In the colormaps in Fig. \ref{fig:teaser}, each color scale was applied to the same underlying dataset but in the experiment, each participant saw 16 colormaps created from a unique pairing of color scales to underlying datasets. That is, they never saw colormaps constructed from the same underlying dataset in more than one color scale. To achieve this experimental design, we produced 256 colormaps (16 color scales $\times$ 16 underlying datasets) and used a Latin square design to balance the assignment of color scales to underlying datasets across participants. This design produced 16 groups, and participants were randomly assigned to group. Within participants, the trials were presented in a randomized block design---participants judged all 16 colormaps in a random order twice within one emotion concept before going on to the next emotion concept (32 trials per block). The order of emotion concepts was also randomized for each participant. 

The following sections describe how we generated the color scales and selected the 16 underlying datasets to construct the colormaps.

\textbf{Generating color scales.}
We created the 16 color scales using \href{https://unc-visualab.org/ColorCrafter}{Color Crafter} \cite{smart2020}, which enables inputing a seed color to generate color scales and provides sliders to shift the overall lightness (L* in CIELCh) and chroma (C* in CIELCh). 

We defined the seed color for each hue in CIE LCh space such that L* = 90, C* = 5, and hue angle (h) was 0, 90, 180, 270 degrees for red, yellow, green, and blue seeds, respectively. We chose these hue angles as approximates of the Hering opponent dimensions (red/green, yellow/blue). We chose these lightness and chroma coordinates after exploring different values in Color Crafter and determining which seeds produced color scales that were roughly monochromatic (single hue). We then converted these coordinates to CIE XYZ values with a white point of CIE Illuminant D65, and converted to RGB values using the \texttt{xyz2RGB} function in MATLAB (assuming a standard sRGB monitor), and then to hexadecimal codes as input seeds for Color Crafter (red: \#ecdfe3 \textcolor[HTML]{ecdfe3}{\rule{1em}{1em}}, yellow: \#e7e2d9 \textcolor[HTML]{e7e2d9}{\rule{1em}{1em}}, green: \#d8e5e2 \textcolor[HTML]{d8e5e2}{\rule{1em}{1em}}, blue: \#dee3ec \textcolor[HTML]{dee3ec}{\rule{1em}{1em}}). Given a seed color, Color Crafter generates nine color scales. We selected the color scale of the target hue that appeared most monochromatic (i.e., had the least hue variability).

Next, we adjusted the chroma and lightness sliders to create four color scales for each hue (Fig. \ref{fig:colorcrafter}). For \textbf{\textit{light high chroma}} color scales, we increased lightness and chroma sliders to their maxima (Fig. \ref{fig:colorcrafter}A). For \textbf{\textit{dark high chroma}} color scales, we kept the chroma slider at its maximum and we reduced the lightness slider to its minimum (Fig. \ref{fig:colorcrafter}B). To create the \textbf{\textit{dark low chroma}} color scales, we kept the lightness slider at its minimum and reduced chroma as far left as possible while preserving monotonicity in the light region of the curve (Fig.  \ref{fig:colorcrafter}D).\footnote{Sliding chroma all the way left (lowering chroma) can cause the chroma of the curve to `dip' in the light range because the color scale crossed zero in the A/B plane of CIELAB space, thereby adding a new hue into the color scale. Aiming to produce roughly monochromatic color scales, we avoided this issue by moving the chroma slider as far left as possible before the `dip'.} 
For \textbf{\textit{light low chroma}} color scales, we kept the chroma slider where it was and minimized the lightness slider (Fig. \ref{fig:colorcrafter}C).

The color coordinates of the color scales are not perfectly controlled (e.g., L* and C* are not matched across the light, high chroma color scales for each hue, see Supplementary Material Figures \ref{fig:color-scale-coordinates} and \ref{fig:color_scale_means}), but we used this approach because Color Crafter produces effective color scales for data visualization in practice \cite{smart2020}. We could have designed tightly controlled color scales, but they would likely have been less effective for visualization, thus weakening the translatability of our results to design practice. Our analyses control for this variation by coding predictors according to the average CIELAB/CIELCh coordinates of the colors scales. See details in Section \ref{sec:Exp1_discrete}.

\textbf{Selecting the underlying datasets.}
We created the colormaps by applying the 16 color scales described above to a global raster dataset of aboveground live woody biomass (AGB) \cite{harris2021}. Each pixel in the dataset has a spatial resolution of 30$\times$30 meters and represents megagrams of AGB per hectare (Mg ha$^{-1}$) in the year 2000. The dataset includes 280 `tiles' across the world (40,000$\times$40,000 pixels per tile). We split each tile into smaller 1,000$\times$1,000 pixel `subtiles' (448,000 subtiles total). We then excluded subtiles that had N/A values, corresponding to 0 Mg ha$^{-1}$, to avoid including subtiles that had easily recognizable geographical features (e.g., rivers or croplands). 

Of the remaining set, we sought to find subtiles that had the most uniform distribution of values, such that when we applied a color scale to produce a colormaps, all colors in the color scale would be well-represented in the map. Although we could have generated synthetic datasets that were perfectly uniform, we used real-world data for ecological validity. We used the following steps to filter the subtiles. First, we rescaled the values from units of biomass density to a standardized scale from 0 to 100. Second, we eliminated subtiles whose medians were outside of a range of 40-60 to avoid skewed distributions. Third, for each of the subtiles we grouped the values into nine bins corresponding to the nine colors within the Color Crafter color scales and computed the pixel count for each bin. We kept subtiles for which the pixel counts of each of the nine bins fell within an acceptable range of deviation from the expected pixel count for each bin under a uniform distribution (1000$\times$1000 pixels/9 bins).\footnote{We operationalized `acceptable' as within ±1/3 of the expected bin counts. We explored other metrics to assess deviation from a uniform distribution, but we found this approach was most effective for ensuring that no bin had an outsized number of pixels.} This method resulted in 44 maps (Fig. \ref{fig:graymaps44}; see CSV file in our \href{https://github.com/SchlossVRL/color_scales_affect}{GitHub} repository for pixel values).

\textbf{Identifying subtiles with similar affective connotation.} Given our aim to test the influence of color on affective connotation of colormaps, we sought to use maps that all had similar affective connotation when presented in grayscale. Thus, any differences observed in our main experiments would likely be due to color and not spatial structure of the maps (e.g., spatial distribution of pixels). We note that such effects were also mitigated by balancing the assignment of color scale to underlying datasets across participants.
 
To find grayscale maps with similar affective connotation, we conducted a pilot study in which 46 participants rated the association between each of the 44 maps (Fig. \ref{fig:graymaps44}) and each of the eight affective terms tested in Exp. 1 (see Section \ref{sec:gray_pilot} for participant demographics). The experimental design and procedure were the same as in Exp. 1, replacing the 16 chromatic colormaps with 44 grayscale colormaps, each rated once.  

To determine the most similar maps in terms of affective connotations, we first correlated the mean associations for each pair of maps across the 8 emotion concepts (averaged over participants), which produced a symmetric 44 $\times$ 44 correlation matrix (Fig. \ref{fig:gray_emo}A). Hierarchical clustering revealed a central cluster of 26 strongly correlated maps (i.e., similar affective connotation). We selected the 16 maps that had the highest mean correlation with each of the other maps. Examining the affective connotation of these 16 maps in grayscale (Fig. \ref{fig:gray_emo}B), they happened to be most associated with \textit{disturbing} and \textit{negative}.  

\textbf{Applying color scales to subtiles.} With the final set of 16 subtiles, we applied the 16 color scales to create the colormaps for this experiment. This process produced 256 colormaps for Exp. 1.  

\textbf{Experiment displays.} 
The experiment was administered in participants' browsers using jsPsych \cite{de2015jspsych, de2023jspsych}. As such, the precise colors each participant saw depended on properties of their own monitors and all reported CIE values are approximations using standard assumptions about RGB displays. This approach is typical in visualization research, which aim to produce results that are robust to variations across displays \cite{stone2014, szafir2018, gramazio2017, mukherjee2022}.

During the experimental task, the participants saw displays each containing one of the colormaps centered on the screen scaled to be 200px × 200px. This size corresponded to 3.8cm × 3.8cm on a display that was 29.5cm × 16.5cm with a resolution of 1920px × 1080px. Below was a slider scale with the endpoints labeled `not at all' and `very much' and an unlabeled midpoint. The background of the monitor was gray (rgb: [128,128,128]).

\subsubsection{Procedure} \label{sec:exp1_procedure}
\begin{figure} [t]
 \centering
 \includegraphics[width=0.7\columnwidth]{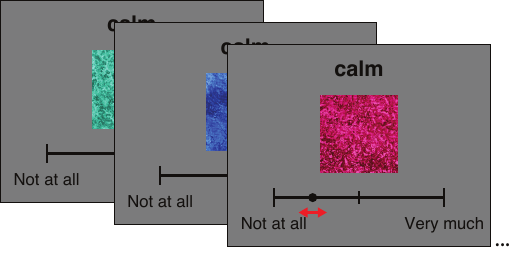}
 \caption{Example trials for rating color-emotion associations for colormaps.}
 \label{fig:colormap_task}
 \vspace{-6mm}
\end{figure}

The participants were instructed that they would be presented with a series of colormaps, one at a time, and their task was to rate how much they associated each map with each of the eight affective concepts on a scale from `not at all' to `very much' (Fig. \ref{fig:colormap_task}). Before beginning the task, they were shown a list of all eight concepts and all 16 colormaps they would be asked to judge during the experiment. The colormaps were shown in two rows of eight maps and were randomly ordered for each participant. Below the maps was an example rating scale. The participants were asked to look at the concepts and maps and consider which map they associated `very much' and `not at all' with each concept. This `anchoring' task was done so participants would know what associating `not at all' and `very much' meant for them for the colors and concepts in the context of this experiment \cite{palmer2013}. The participants were also asked to use the full range of the scale when making their ratings and were instructed that if they somewhat associated a map with a concept, then they should click partway between the middle point and right side of the scale. 
 Participants were informed that they would be asked to rate each map for a given concept before moving on to the next concept. They were also told to make ratings based on their first intuition because we were interested in their initial impressions of each map for a given concept.

The participants were then instructed on how to use the response scale by sliding the cursor to the position where they wanted to make their rating and then letting it go to record their response. They were given four training trials, asking them to move the slider to each endpoint and halfway between the center and each endpoint. If they did not place the slider near the instructed location, they were asked to try again until the correct location was selected. Before each block, participants were told the concept they would be asked to judge for that block. They were then presented with the 32 trials for that block, each separated by 250ms inter-trial interval. 

\subsection{Results and Discussion}
\begin{figure}[t]
 \centering
 \includegraphics[width=0.85\columnwidth]{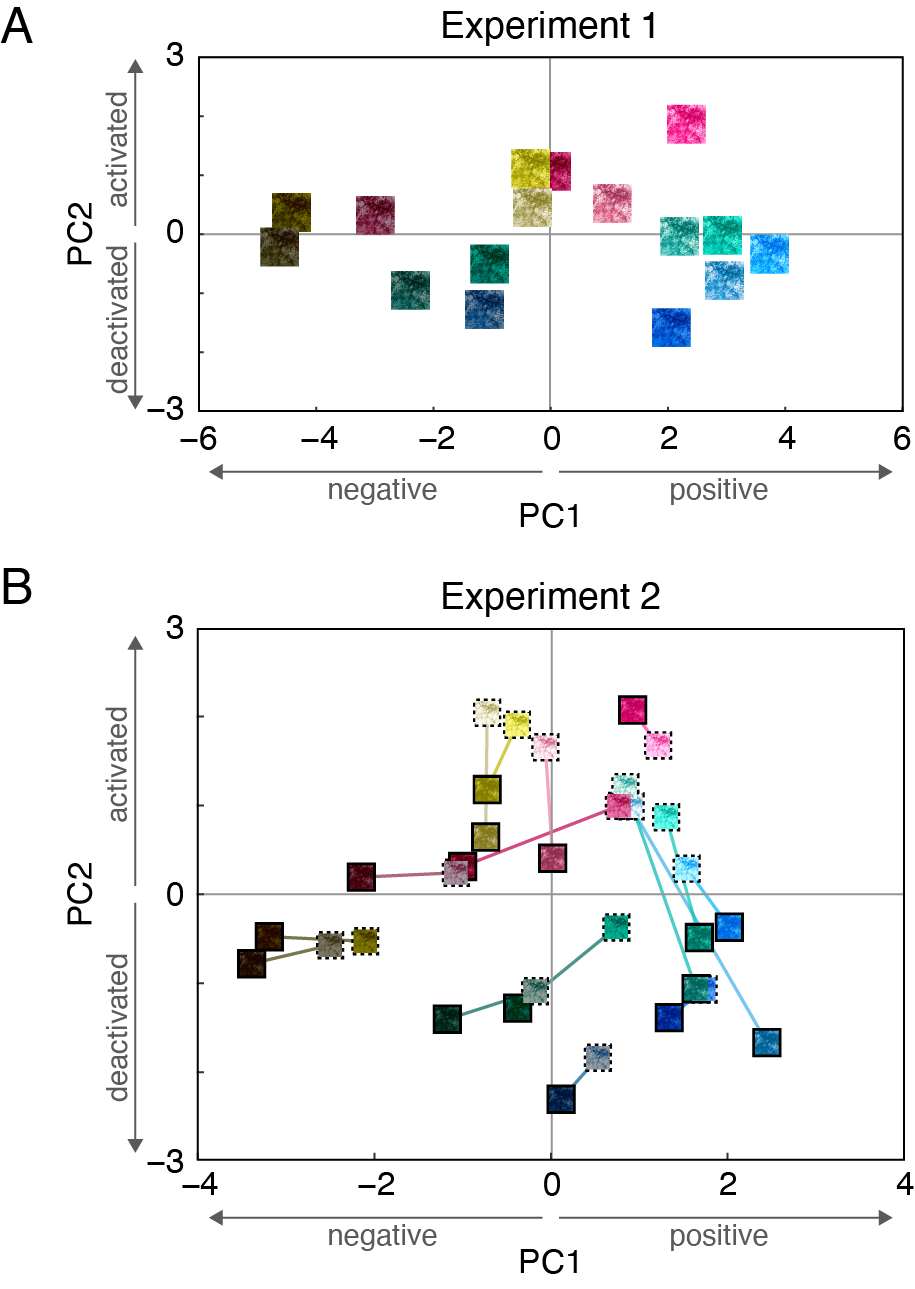}
 \caption{Results of the PCA conducted on association data from (A) Exp. 1 and (B) Exp. 2. In (A), each of the 16 colormaps represents all colormaps produced using the color scale shown in that map. In (B), the 16 colormaps with solid/dashed outlines represent dark/light shifted colormaps, respectively. Lines connect dark and light shifted colormaps generated from the same color scale. }
 \label{fig:exp1and2_pca}
 \vspace{-4mm}
\end{figure}

We first analyzed the results in terms of discrete emotions (Fig. \ref{fig:exp1_assoc}) and then examined the underlying dimensional structure (Fig. \ref{fig:exp1and2_pca}A). 
All analyses were conducted using R version 4.2.1 \cite{r2013r}.

\subsubsection{Analysis of discrete emotions} \label{sec:Exp1_discrete}
\begin{figure}[t]
 \centering
 \includegraphics[width=1.0\columnwidth]{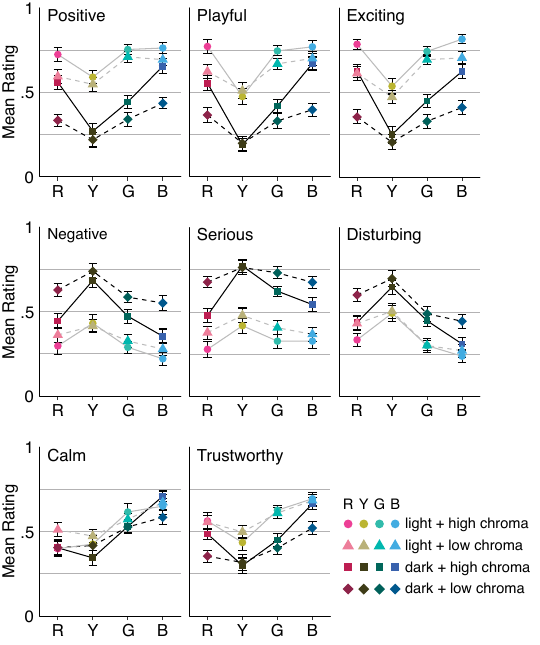}
 \caption{Mean association ratings for colormaps created from each of the 16 color scales and the eight affective concepts. The x-axis represents hue (\textbf{R}ed, \textbf{Y}ellow, \textbf{G}reen, \textbf{B}lue) and the separate lines represent lightness (dark: dark gray, light: light gray) and chroma (high chroma: solid, low chroma: dashed). The mark color represents the middle color of the color scale used to construct the corresponding colormaps. Error bars represent +/- standard errors of the means (SEM).}
 \label{fig:exp1_assoc}
 \vspace{-6mm}
\end{figure}

Fig. \ref{fig:exp1_assoc} shows the mean association ratings for each of the eight emotion concepts and each of the 16 color scales. We analyzed the pattern of data for each concept using linear mixed-effect regression (LMER) models \cite{lme42015, baayen2008mixed}, analogous to the models in \cite{schloss2020}, to predict participants' association ratings between the colormaps and that concept. 
We included fixed effects for each color scale's mean lightness (L), chroma (C), a* (red/green), and b* (yellow/blue) from CIELAB and CIELCh color spaces.
We also included interactions for each opponent hue pair (L $\times$ C $\times$ a* and L $\times$ C $\times$ b*) and pairwise interactions therein following \cite{schloss2020}, and included by-subject random slopes and intercepts for each fixed effect. We note that averaging a* or b* would be nonsensical if the color scales were not roughly monochromatic (Fig. \ref{fig:color-scale-coordinates}) (e.g., averaging opposite hues along a* or b* would cancel to gray). We $z$-scored each predictor to put them on the same scale so that the resulting model weights could be comparable across predictors.

Given that we fit eight models, one for each emotion, we applied a Bonferroni correction at the model level to correct for multiple comparisons (critical $\alpha$ of $.05/8 = .00625$). 
Fig. \ref{fig:dots}A highlights key findings and the full model results are shown in Supplementary Material Table \ref{tab:exp1-lmer-results-2col}. The dots in Fig. \ref{fig:dots}A show the sign (+/-) and magnitude (diameter) of the beta weights for each effect that was significant after correction (\textit{ns} indicates not significant). None of the 3-way interactions were significant so they were excluded from Fig. \ref{fig:dots}A.

\begin{figure}[ht!!!]
 \centering
 \includegraphics[width=0.85\columnwidth]{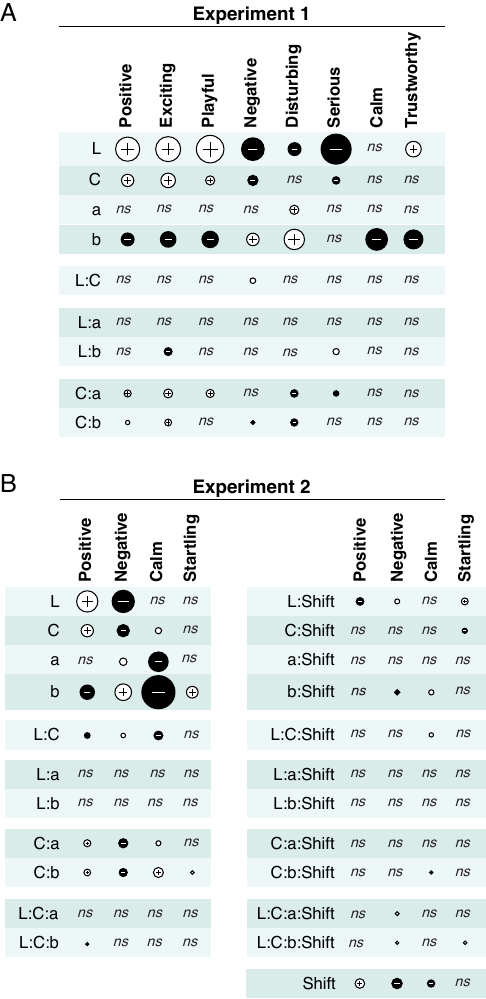}
 \caption{Summary of LMER model results for (A) Exp. 1 and (B) Exp. 2 (see Supplementary Material Tables \ref{tab:exp1-lmer-results-2col} and \ref{tab:exp2-lmer-results-2col} for the full output). The factors included lightness (L*), chroma (C*), red/green (a*), yellow/blue (b*), and interactions therein. Dots with $+$/$-$ symbols indicate significant effects with positive/negative beta weights, respectively (`\textit{ns}' indicates not significant). Dot diameter is proportional to beta weight to represent effect size (diameter minimum was set to 1.5 pixels to ensure dot visibility).}
 \label{fig:dots}
 \vspace{-6mm}
\end{figure}

Generally, the pattern of associations for \textit{playful} and \textit{exciting} were similar to \textit{positive} (Fig. \ref{fig:exp1_assoc}, row 1), being strongly associated with lighter, higher chroma, and bluer colors. These emotions also all shared a chroma $\times$ red/green interaction--- higher chroma colormaps were especially more \textit{positive}, \textit{playful}, and \textit{exciting} for redder colormaps than greener colormaps. However, the three emotions differed in terms of their other interactions. For \textit{positive} and \textit{exciting}, a chroma $\times$ yellow/blue interaction indicated that higher chroma colors were especially more \textit{positive} and \textit{exciting} when the colors were bluer. For \textit{exciting}, lightness interacted with yellow/blue, indicating that darker colors were especially less exciting when they were yellower. 

The pattern of associations was also similar among \textit{negative}, \textit{serious}, and \textit{disturbing} (Fig. \ref{fig:exp1_assoc}, row 2), being associated with darker, lower chroma colors. However, whereas \textit{negative} was associated with yellower colors with no effect of red/green, \textit{serious} had no effect of yellow/blue, and \textit{disturbing} was associated with redder colors. These three emotions also differed in terms of which interactions were significant. For \textit{negative} only, a chroma $\times$ lightness interaction indicated that the degree that low chroma colors were more negative was greater when the colormaps were darker. For \textit{negative} and \textit{disturbing}, a chroma $\times$ yellow/blue interaction indicated that increasing chroma especially reduced negativity/disturbingness when colors were bluer. For \textit{serious} and \textit{disturbing}, chroma interacted with red/green, though the patterns were different. For \textit{serious}, greener colors were especially more serious when chroma was higher, whereas for \textit{disturbing}, redder colors were especially more disturbing when chroma was lower. 

Finally, \textit{trustworthy} was similar to \textit{calm} (Fig. \ref{fig:exp1_assoc}, row 3) in that both were associated with bluer colors with no effect of chroma, but they differed in that \textit{trustworthy} was associated with higher lightness. 

Together, these results answer our core research question: it is indeed possible to produce colormap data visualizations with strong lightness variation that also have systematic affective connotation.

\subsubsection{Analysis of underlying dimensional structure} 
To understand the nature of the underlying color-affect space, we conducted principal component analysis (PCA) on the mean ratings for colormaps constructed from each color scale for each affective term. This space is well-characterized by a 2D solution, illustrated in Fig. \ref{fig:exp1and2_pca}A. Each of the 16 points in this space represents each of the 16 color scales (combined over colormaps produced from that color scale). The color scale loadings on PC1 were positively correlated with how \textit{positive}, \textit{playful}, and \textit{exciting} the colors were ($r(14) = 0.98, 0.98, 0.97$, all $p$s $< 0.001$, respectively) and negatively correlated with how \textit{negative}, \textit{serious}, and \textit{disturbing} the colors were ($r(14) = -0.99, -0.91, -0.97$,  all $p$s $< 0.001$, respectively). As such, PC1 seems to represent \textit{valence} \cite{posner2005} so we labeled it positive vs. negative. PC2 was most strongly correlated with how \textit{calm} the colors were ($r(14) = -0.73$, $p < 0.001$) and thus seems to represent arousal/activation \cite{posner2005} so we labeled it activated vs. deactivated. The full set of correlations are in Supplementary Material Fig. \ref{fig:emotion-PC-corr}. The dimensions of this affective space are similar to dimensions for single colors \cite{palmer2013, valdez1994, adams1973, jonauskaite2025} and color palettes \cite{bartram2017}.\footnote{The dimensions pleasure, arousal, and dominance (PAD) for emotion are analogous to evaluation, potency, and activity (EPA) for semantic associations \cite{valdez1994}. These citations include studies reporting PA or EA as key dimensions.} 

To characterize the color variation along each PC, we used multiple linear regression with color scale coordinates along the PC as a dependent measure and mean L*, C*, a*, and b* values of each scale as predictors. PC1 (valence) varied from darker, lower chroma, and yellower (negative) to lighter, higher chroma, and bluer (positive) (adjusted model $R^2$ = 0.924). PC2 (activation) varied from greener, bluer, and darker (less activated) to redder, yellower, and lighter (more activated) (adjusted model $R^2$ = 0.836). The full set of results are in Supplementary Material Table \ref{tab:exp1-pc-regression}.

\textbf{Summary.} Exp. 1 showed that colormaps can have clear affective connotation while preserving the lightness contrast that is important for fine-detail spatial vision. These associations were primarily driven by overall lightness, as well as chroma and yellowness/blueness of the colormaps. Analysis of emotion dimensions showed that higher valence corresponded to lighter, higher chroma, and bluer colormaps, and higher activation corresponded to lighter, yellower, and redder colormaps. Designers can leverage our findings for discrete emotions and reference our affective space when selecting color scales to convey desired emotions for colormap data visualizations. 

%%%%%%%%%%%%%%%%%%%%% EXPERIMENT 2 %%%%%%%%%%%%%%%
\section{Experiment 2}
In Exp. 1, the values of the subtiles were well-distributed over the range, which produced colormaps for which all colors in the color scale were well-represented. However, in real-world maps, values can be biased toward one endpoint of the range (e.g., larger or smaller values), resulting in colors that are biased toward one end of the color scale (e.g., darker or lighter colors). Exp. 2 investigated whether affective connotation of colormaps depends simply on the color scales used to construct the colormaps (null; data-agnostic hypothesis), or rather depends on the proportion in which each color appears in the colormap, determined by the underlying dataset (data-dependence hypothesis). 

\subsection{Methods}

\subsubsection{Participants}
The participants ($n=34$) were UW--Madison undergraduates who received extra credit in their Psychology course (mean age: 18.9, range: 18--21). They provided demographic information in free-response text boxes. Their genders included 27 women and 7 men (no other genders reported). Their race/ethnicity included 24 White, 5 Asian, 4 Hispanic, 2 Latino/Latina, 1 Middle Eastern, 1 Native American, and 1 Hmong (the number of entries exceeded the number of participants because some participants indicated $>1$ race/ethnicity). 
We excluded 1 participant who indicated atypical color vision (assessed as in Exp. 1).

\subsubsection{Design, displays, and procedure}

\begin{figure}[t]
 \centering
 \includegraphics[width=0.9\columnwidth]{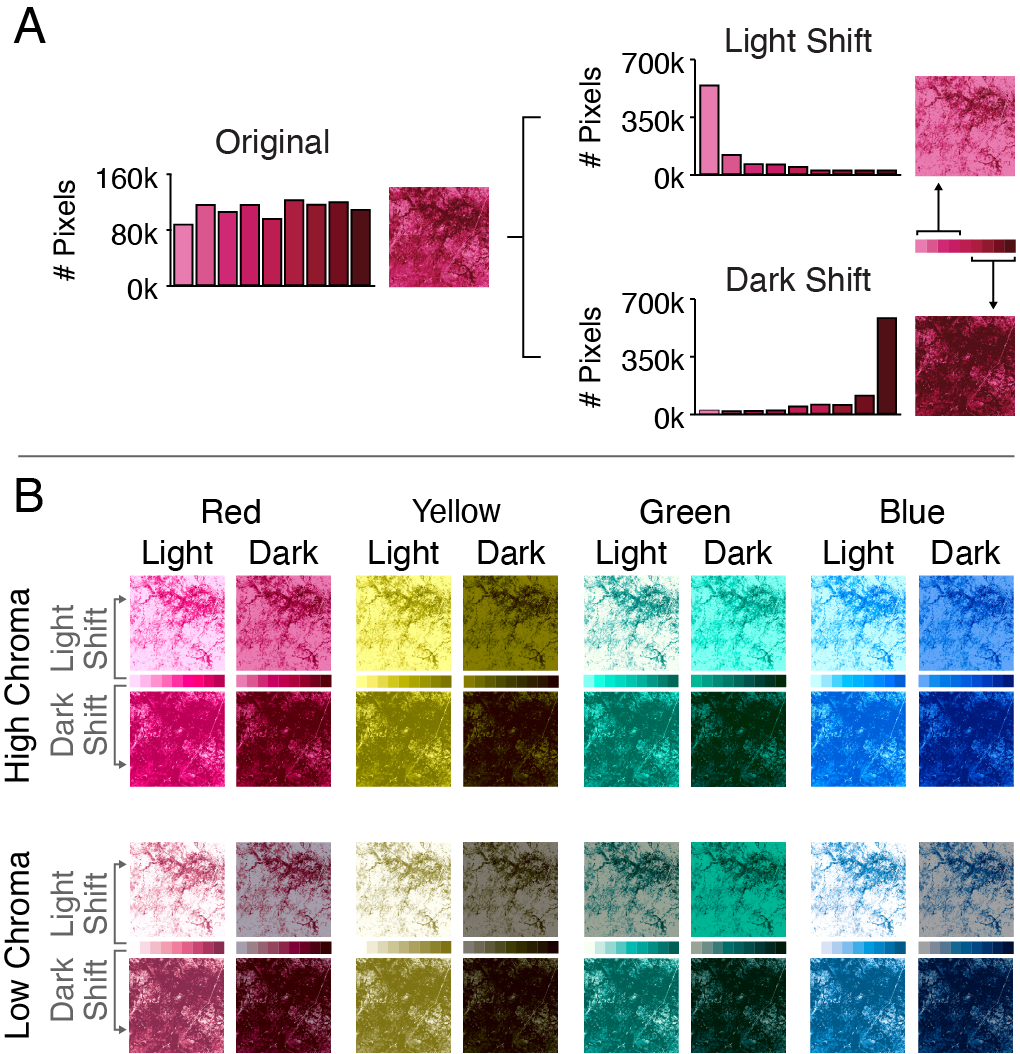}
 \caption{(A) Diagram showing how the underlying dataset for a colormap from Exp. 1 was light shifted and dark shifted to produce the colormaps in Exp. 2. The histograms for each colormap show the number of pixels in each color from the color scale used to produce the colormap (the bar colors correspond to pixel colors in the colormap). (B) Example colormaps that were created from the same 16 color scales from Exp. 1 (Fig. \ref{fig:teaser}) applied to light shifted or dark shifted datasets.}
 \label{fig:Exp2_shift}
 \vspace{-6mm}
\end{figure}

The methods were the same as Exp. 1 with the following exceptions:

\textbf{Shifted colormaps}. Participants judged two versions of colormaps generated from each color scale, a light and dark shifted version (16 color scales $\times$ 2 shift conditions = 32 maps). Fig. \ref{fig:Exp2_shift}A illustrates light and dark shifting, starting from an original colormap from Exp 1. The histogram next to each colormap shows the number of pixels mapped to each color of the color scale (bar color corresponds to color in the color scale). In the light shifted maps, the majority of pixels are in the range of values corresponding to light colors in the color scale, whereas the opposite is true in the dark shifted maps. 

To create the shifted versions, we first normalized the values in each subtile (0--1; min-max normalization). Next, we created `light shifted' maps (i.e., a greater proportion of lighter values) by applying the following formula to each pixel $x$ in the data matrix

$$
b - \frac{(b - x)^p}{(b - a)^{p - 1}}
$$

where $a$ and $b$ were the limits of allowable data values (0 and 1, respectively). The parameter $p$ determined the degree of shift (larger values of $p$ resulting in larger shifts). 

The analogous equation for making `dark shifted' colormaps was  

$$
\frac{(x - a)^{p}}{(b - a)^{p - 1}} + a
,$$

We set the value of $p$ to be 4 after visually inspecting the light and dark shifted maps for a range of parameter values. This parameter value resulted in shifted maps that were perceptually distinct from the non-shifted version while preserving enough visual detail that they still appeared to be generated from the same data distribution.

Fig. \ref{fig:Exp2_shift}B shows light and dark shifted maps for each of the 16 color scales.
For a given participant, the light and dark shifted maps for a given color scale started from the same underlying dataset, and each color scale had a unique dataset (as in Exp. 1).

\begin{figure}[ht!]
 \centering
 \includegraphics[width=0.8\columnwidth]{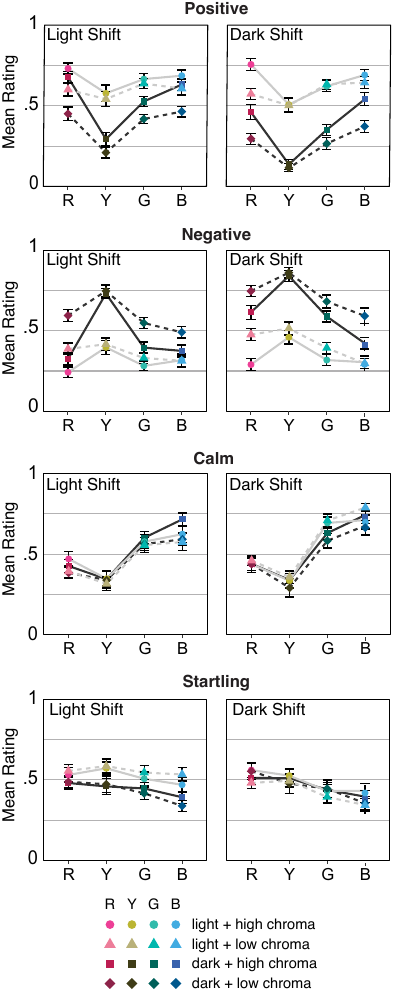}
 \caption{Mean color-concept association ratings for light shifted (left) and dark shifted (right) colormaps from Exp. 2. Data are plotted in the same manner as Fig. \ref{fig:exp1_assoc} from Exp. 1.}
 \label{fig:Exp2_assoc}
 \vspace{-7mm}
\end{figure}

\textbf{Emotion concepts.} We reduced the set of emotion concepts because associations were highly correlated among concepts in Exp. 1 (Fig. \ref{fig:concept-concept-corr}). We retained \textit{positive} and \textit{negative} to span valence and \textit{calm} as an endpoint of activation. Typically, exciting is considered opposite of calm \cite{bartram2017, posner2005} but that was not the case in Exp. 1  (Supplementary Material Figs. \ref{fig:concept-concept-corr} and \ref{fig:emotion-concept-PCA}). Thus, we sought a different opposite of \textit{calm} using a natural language processing (NLP) word embedding model (GloVe \cite{pennington2014glove}) and arrived at \textit{startling} (see  Supplementary Material Section \ref{sec:startling} for details). 
 
 During the experiment, the participants rated their association between each of the 32 colormaps twice for each of the four emotions in a randomized block design as in Exp. 1 (256 trials in total). In the task instructions, we showed participants the full set of concepts for anchoring (including those from Exp. 1) so they would have the same affective context for performing the task. But, during the task, they only judged \textit{positive}, \textit{negative}, \textit{calm}, and \textit{startling}.

\subsection{Results and Discussion}
As in Exp. 1, we analyzed the results in terms of discrete emotions (Fig. \ref{fig:Exp2_assoc}) and then examined the underlying dimensional structure (Fig. \ref{fig:exp1and2_pca}B). 

\subsubsection{Analysis of discrete emotions} 
Fig. \ref{fig:Exp2_assoc} shows the mean association ratings for each of the four affective concepts and each of the 16 color scales, depending on whether the underlying datasets were light or dark shifted. 

We first analyzed the data using the same LMER models as Exp. 1 but included shift as a new fixed factor (+1 = light shift, -1 = dark shift) and tested for interactions between shift and the other factors. We corrected for multiple comparisons at the model level as in Exp. 1, but here there are four models instead of eight (critical $\alpha$ of $0.05/4 = 0.0125$). Also here, effects combined over shift included 4 judgments (2 shift levels $\times$ 2 repetitions), as opposed to 2 judgments in Exp. 1.  With a less strict $\alpha$ and more data per color scale to produce more stable measures, some factors that did not reach significance in Exp. 1 may be considered significant here. The full output of the models are shown in Supplementary Material Table \ref{tab:exp2-lmer-results-2col} and key results are summarized in Fig. \ref{fig:dots}B. The effects of color dimensions combined over shift are shown in Fig. \ref{fig:dots}B (left), and can be compared to Exp. 1 (Fig. \ref{fig:dots}A). The effects involving shift are in Fig. \ref{fig:dots}B (right). We first compare the results combined over shift to the results of Exp. 1, and then focus on effects involving shift to test the data-dependence hypothesis.

\textbf{Effects combined over shift conditions.}  As in Exp. 1, \textit{positive} was associated with lighter, higher chroma, and bluer colormaps, and chroma interacted with red/green and yellow/blue. Unlike Exp. 1, lightness $\times$ chroma and lightness $\times$ chroma $\times$ yellow/blue interactions indicated that the increased association for higher chroma colors with \textit{positive} was more extreme when colors were darker, especially when they were also bluer (Fig. \ref{fig:Exp2_assoc}). Although these effects were not significant in Exp. 1, the pattern was similar (Fig. \ref{fig:exp1_assoc}) and the weights were in the same direction (Table \ref{tab:exp1-lmer-results-2col}). These effects are also analogous to previous results for single colors when rated in terms of \textit{happiness} \cite{schloss2020}. 

Also as in Exp. 1, \textit{negative} was associated with darker, lower chroma, yellower colormaps, and increasing chroma especially reduced negativity when colors were bluer and darker. Unlike Exp. 1, an effect of red/green and a red/green $\times$ chroma interaction indicated redder colors were more negative, particularly when chroma was low. Again, although these effects were not significant in Exp. 1, the weights were in the same direction (Table \ref{tab:exp1-lmer-results-2col}). 

Finally, like Exp. 1, \textit{calm} was associated with bluer colormaps, but unlike Exp. 1 it was associated with greener, higher chroma colormaps. The weight for red/green was in the same direction as Exp. 1 but not for chroma (Table \ref{tab:exp1-lmer-results-2col}), and the reason is unclear. Moreover, in Exp. 2 chroma interacted with lightness, red/green, blue/yellow, but these interactions are difficult to interpret as the curves for \textit{calm} largely overlap in Fig. \ref{fig:Exp2_assoc}. New to Exp. 2, \textit{startling} was associated with yellower colormaps, suggesting it is somewhat opposite of \textit{calm}. 

Overall, the mean associations between matched color scales and emotions across Exp. 1 and 2 (averaged over shift condition) were strongly correlated (\textit{positive}: $r = 0.97, p < 0.001 $, \textit{negative}: $r = 0.98, p < 0.001 $, and \textit{calm}: $r = 0.90, p < 0.001 $). The remaining analyses separated the data by shift to test the data-dependence hypothesis. 

\textbf{Effects of shift.} Overall, light shifted colormaps were more associated with \textit{positive} and dark shifted colormaps were more associated with \textit{negative} (Figs. \ref{fig:dots}B and \ref{fig:Exp2_assoc}). Shift $\times$ lightness interactions for both concepts indicated that the effect of color scale lightness was greater for dark shifted maps (i.e.,  dark shifted colormaps constructed from darker color scales were especially less \textit{positive} and more \textit{negative)}. Dark shifting also amplified the extent to which yellower colormaps were more \textit{negative} than bluer colormaps (shift $\times$ yellow/blue interaction), especially for darker, higher chroma colors (shift $\times$ lightness $\times$ chroma  $\times$  yellow/blue interaction). \textit{Negative} also had a 4-way interaction with red/green, such that dark shifting magnified differences due to lightness and chroma for redder colormaps, which were less extreme for greener colormaps (Fig. \ref{fig:Exp2_assoc}). Together, these results contribute to the narrative that effects of color dimensions on \textit{positive}/\textit{negative} associations are generally more extreme when colors are darker, as determined by darker colors scales or dark shifting the underlying dataset. 

Turning to the other two emotions, dark shifting overall increased the associations with \textit{calm}, especially for colormaps that were bluer (shift $\times$ yellow/blue interaction). Fig. \ref{fig:dots}B shows additional higher order interactions for \textit{calm}, but they are difficult to interpret because of the overlapping pattern of associations (Fig. \ref{fig:Exp2_assoc}). For \textit{startling}, shift had no overall effect, but it interacted with lightness---increasing lightness made colormaps especially more startling when they were light shifted.

The effects of shift reported in this section support the data-dependence hypothesis, suggesting that modeling affective connotation of colormaps based only on colors in the color scale is insufficient. It is necessary to account for the frequency colors appear in the colormap as determined by the underlying data. To test this implication directly, we compared two mixed-effect regression models for each emotion. The `data-agnostic' models predicted association ratings using the mean L*, C*, a*, and b* coordinates of the color scales used to construct the colormaps (same as the models in Exp. 1). The `data-aware' models used the average L*, C*, a*, and b* coordinates of the pixels in each colormap, which weighted the colors in the color scales by the frequency they appeared in the colormaps.
 To enable model comparisons, we fit these models using maximum likelihood (ML) as opposed to restricted maximum likelihood (REML) as the objective function.  Table \ref{tab:exp2_model_comparisons} shows that the data-aware models outperformed the data-agnostic models for all four emotions according to AIC ($\downarrow$ better), BIC ($\downarrow$ better), and conditional $R^2$ ($\uparrow$ better) metrics.  Individual model results are summarized in Tables \ref{tab:exp2-scales-lmer-results-2col} and \ref{tab:exp2-maps-lmer-results-2col}

\begin{table}[H]
\small	

    \centering
    \caption{Model comparisons between mixed effects models using color scale and colormap predictors (\textit{df} = 18 for all models).}
    \begin{tabular}{llccc}
    \hline
         \textbf{Emotion} & \textbf{Model Type} & \textbf{AIC} & \textbf{BIC} & \textbf{Cond.} \boldsymbol{$R^2$}  \\
             \hline
         Positive   & Data-agnostic  & -441.05 & -338.72 & 0.518\\
                    & Data-aware  &-566.66 & -464.33 & 0.558\\
            \hline
         Negative   & Data-agnostic  &-421.51 & -319.17 & 0.538 \\
                    & Data-aware &-628.47 & -526.14 & 0.593 \\
             \hline
         Calm       & Data-agnostic  &-130.48 &  -28.14 & 0.530\\
                    & Data-aware  & -361.37 & -259.03 & 0.593\\
             \hline
         Startling  & Data-agnostic  &-305.99 & -203.66 & 0.489\\
                    & Data-aware  &-395.13 & -296.60 & 0.534\\
         \hline
    \end{tabular}

    \label{tab:exp2_model_comparisons}
     \vspace{-3mm}
\end{table}

\subsubsection{Analysis of underlying dimensional structure} 
Fig. \ref{fig:exp1and2_pca}B shows the results of a PCA on the mean association ratings from Exp. 2. Each of the 32 points in this space represents the mean of all colormaps constructed from each color scale $\times$ shift level. Light shifted and dark shifted colormaps produced using the same color scale are connected by a line colored according to the middle color of the color scale. If shift had no effect, then colormaps produced using the same color scale would share the same position in the space. However, that was not the case, and some colormaps even traversed different quadrants of the space. For example, colormaps produced from the dark, high chroma red color scale (Fig. \ref{fig:Exp2_shift}A) were positive when light shifted and became negative when dark shifted (Fig. \ref{fig:exp1and2_pca}B). 

The dimensions of this affective space were well mapped to valence (positive vs. negative) and arousal/activation (activated vs. deactivated) as in Exp. 1.; see Supplementary Material Fig. \ref{fig:emotion-PC-corr} for correlations. 
Using multiple linear regression, we characterized the color variation along each PC (see Table \ref{tab:exp1-pc-regression}), but instead of data-agnostic predictors we used data-aware predictors, shown to fit the data better (Table \ref{tab:exp2_model_comparisons}). The PCs had similar structure as in Exp 1---PC1 (valence) increased with lightness, chroma, and blueness (adjusted model $R^2$ = 0.716) and PC2 (activation) increased with lightness, redness, and yellowness (adjusted model $R^2$ = 0.832). The one difference is valence was not related to red/green in Exp. 1 but increased with greenness in Exp. 2. 

\textbf{Summary.} Supporting the data-dependence hypothesis,  affective connotation depended not only on the color scales used to construct colormaps, but also the proportion that the colors appear in the colormap, as determined by the underlying dataset. The colors that were more prevalent contributed more to the affective connotation of colormaps. These results imply that for a designer to anticipate the affective connotation of colormaps, it is important that they use a data-aware approach that accounts for how the colors they choose will appear in the context of the visualizations, given the relative frequency of values in the data.

\section{General Discussion}
Our study was motivated by the premises that (a) affective connotations of colors are largely driven by lightness level, and (b) for colormap data visualizations to reveal fine details in spatial structure, they need to vary in lightness. Our primary aim was to investigate whether colormaps that vary in lightness to support spatial vision can communicate distinct affective connotations. Indeed, we found systematic color-emotion associations, similar to previous findings for single colors \cite{schloss2020, dandrade1974, wright1962, jonauskaite2019mood, dael2016}. For example, positive emotions were associated with lighter, higher chroma, and bluer colormaps, whereas negative emotions had the opposite associations. And, similar to previous results for color palettes \cite{bartram2017}, cooler colormaps were associated with \textit{calm}. 

However, we also found differences from previous work on color palettes. For example, whereas yellower color palettes were \textit{positive}, \textit{playful}, and \textit{exciting} \cite{bartram2017}, we found that those emotions were associated with bluer colormaps. This yellow/blue discrepancy can be understood in terms of the role of chroma and lightness. 
In Bartram et al.'s \textit{exciting}, \textit{positive}, and \textit{playful} color palettes, the warm colors were all relatively light and high in chroma, but for the present colormaps, the colors varied widely in lightness and chroma (Fig. \ref{fig:teaser}). As yellower hues darken they can no longer be high in chroma \cite{wyszecki1982, munsell1921}, and a decrease in both lightness and chroma makes them especially negative \cite{schloss2020, jonauskaite2025}. In contrast, as bluer hues darken, they can become higher in chroma \cite{wyszecki1982, munsell1921}, and increasing chroma as blues darken helps preserve the positive association \cite{schloss2020}. It follows that for colormaps varying in lightness, bluer colormaps should be more positive than yellower colormaps because they shift less toward negative as the colors get darker. These results imply that a polychromatic color scale like \texttt{viridis}, which has light, high chroma yellows and dark, high chroma blues, would be strongly positive. Indeed, evidence suggests colormaps produced with \texttt{viridis} evoke positive emotional responses \cite{zhang2021}. 

Our study also aimed to test the data-dependence hypothesis, proposing that affective connotation depends not only \textit{which} colors are present in the visualization, but also \textit{how much} those colors appeared in the image as determined by the underlying dataset. Supporting this hypothesis, Exp. 2 showed that shifting the underlying dataset modulated color-emotion associations (e.g., light shifted colormaps were associated more with \textit{positive} and less with \textit{negative} or \textit{calm} than dark shifted maps). Moreover, the pattern of associations was better predicted by data-aware models that accounted for the color distributions of pixels within the colormap images compared to data-agnostic models that gave equal weight to every color in the color scale.  

In the present study, shifting the underlying dataset primarily modulated the lightness of pixels in the maps because the color scales were roughly monochromatic and most of them varied in chroma in an inverted U function (the high chroma colors were in the middle of the scale; Fig. \ref{fig:color-scale-coordinates}). However, if a polychromatic color scale was used (e.g., \texttt{viridis}), shift would influence hue representation as well as lightness. Moreover, data can have all kinds of distributions that influence which colors are represented in the colormaps. For example, if we had shifted the data to the mid-range, the colors would have had mid-level lightness and would have been higher in chroma for most color scales in our set. 

These points emphasize the importance of data-aware visualization design, which accounts for not only which design features encode data (e.g., colors, shapes, textures), but also how those design features are instantiated in a visualization, given the numeric and semantic properties of the underlying dataset. A data-aware approach for designing colormaps to have particular affective connotations might look like this: (a) identify colors of the desired affective connotation, (b) examine the histogram of the dataset, and (c) apply a color scale such that the colors identified in (a) get mapped to the most frequent colors within the dataset. Other examples of data-aware colormap design involve (a) selecting/modifying color scales so they help observers perform specific kinds of tasks or highlight what Zeng et al. \cite{zeng2021} call `meaningful features' of data \cite{zeng2021, samsel2018ColorMoves, rogowitz1996, tominski2008} and (b) examining the spatial structure of data \cite{zimnicki2023, sibrel2020} and the concepts that the data represent \cite{soto2023, schoenlein2023} to anticipate whether observers will infer that darker or lighter colors represent larger quantities in the colormap. Here, our emphasis is on colormaps, but questions remain concerning the importance of data-awareness for categorical visualizations. Blair et al.\cite{blair2025} suggested links between color and emotion for categorical visualizations are similar to those of isolated color palettes, implying data agnosticity. But, data-awareness may become more important for categorical visualizations if there is greater discrepancy in the area occupied by each color (e.g., different sized bars in bar charts or regions in choropleth maps).

Our study raises additional open questions for future work. First, how do color and spatial factors combine to influence affective connotations of colormaps? Here, we controlled the affective connotation of spatial structure by choosing maps that had the most similar affective connotation in grayscale. Future work can test whether spatial and color factors influence affective connotation independently or interact in interesting ways. A second question concerns potential implications of our results for judgment and decision-making pertaining to visualizations. Anderson and Robinson \cite{anderson2022} reported that congruence between visualization topic and colors increased judgments of appropriateness. Do such design factors influence down-stream decisions (e.g., likeliness to support a cause) or the interpretations of data in storytelling contexts \cite{boy2017, liem2020}? Finally, our study was conducted in the US among college undergraduates, but how might these results vary across cultures? Although studies have found cross-cultural commonalities in color-emotion associations, there are also systematic cultural differences \cite{jonauskaite2025, dandrade1974, palmer2013, tham2020systematic, ou2004_single, jonauskaite2019machine, xin2004, adams1973, jonauskaite2020universal}. For example, Xin et. al. \cite{xin2004} found that color-emotion associations for participants from Thailand were especially driven by lightness, compared to those from Hong Kong and Japan. A comprehensive understanding of affective connotation of color in visualizations will require a cross-cultural approach. 

\section{Conclusion}
We found that colormap visualizations that vary in lightness to support fine-detail spatial vision can communicate clear affective connotations.  Affective connotation depends not only on which colors appear in a colormap but also on color frequencies as determined by the underlying dataset, emphasizing the importance of data-aware design. This study was a step toward understanding affective connotation in colormap design, and future work can extend this approach  by studying data-aware visualization design involving polychromatic color scales, interactions between color and spatial factors, and variations across populations.

\acknowledgments{%
We thank Danielle Szafir, Psyche Loui, Domicele Jonauskaite, and Melissa Schoenlein. This work was supported in part by NSF BCS-1945303. Any opinions, findings, and conclusions or recommendations expressed in this material are those of the authors and do not necessarily reflect the views of the NSF. Use of AI: Github Copilot helped write code for experiment tasks and Claude helped format tables in LaTeX.
}

 \clearpage
\bibliographystyle{abbrv-doi-hyperref-narrow}

\bibliography{template}

\clearpage
\appendix % You can use the `hideappendix` class option to skip everything after \appendix

\renewcommand*{\thesection}{S}
\counterwithin{figure}{section}
\counterwithin{table}{section}
%\onecolumn

% \balance

\section{Supplementary Material}\label{sec:supplementary}

\subsection{Participants in pilot study on finding gray maps with similar affective connotation} \label{sec:gray_pilot}
The participants were 46 undergraduates from UW--Madison who received extra credit in their Psychology course (mean age: 18.96, range: 18 to 23). They provided their demographic information in free-response text boxes. The genders of participants included 39 women, 8 men, and 1 non-binary person. Their race/ethnicity included 39 White, 3 Asian, 1 Native American, 1 Mixed, 1 African-American, 1 Dominican, and 1 Indian (please note that the number of entries was greater than the number of participants because some participants indicated more than one race/ethnicity). Participants completed this experiment on their own devices.

\subsection{Finding a semantic opposite of \textit{calm}} \label{sec:startling}
Our approach was based on using semantic embeddings derived from GloVe \cite{pennington2014glove}, a natural language processing (NLP) word embedding model. GloVe has been validated not only as a good NLP model but also as a good proxy for people's semantic representations as measured using behavioral and neural measurements \cite{pereira2018toward, grand2022semantic}, making it a suitable model for our purposes.
Since the dimensions of GloVe need not perfectly align with the dimensions latent in people's association ratings from Exp. 1, we needed a method to map the human principal component space to GloVe space. 
To do so, we estimated a transformation matrix $M$ that mapped from the 300-D GloVe space to PC space using scipy's \texttt{lstq} least-squares solver. 
Using $M$, any word's GloVe embeddings could be mapped to human PC space.
We began by finding the words most dissimilar to \textit{calm} based on their GloVe embeddings, projecting them into PC space using $M$ and then visualizing the position of each concept. 
We iteratively performed this process with candidate antonyms, until we found a word that loaded on the opposite end of the PC that calm loaded positively on \textit{without} loading much on either direction on the \textit{positive} -- \textit{negative} PC. 
We arrived at \textit{startling} as a good candidate based on this approach.

\begin{table*}[ht!]
    \centering
    \caption{Hexadecimal codes for each color within the 16 color scales used in Experiments 1 and 2 (indicated by overall hue (\underline{R}ed, \underline{Y}ellow, \underline{G}reen, \underline{B}lue), lightness (\underline{L}ight, \underline{D}ark), and chroma (\underline{Hi}gh, \underline{Lo}w).}
    \begin{tabular}{lll*{9}{c}}
    \toprule
    Hue & Lightness & Chroma & color 1 & color 2 & color 3 & color 4 & color 5 & color 6 & color 7 & color 8 & color 9 \\
    \midrule
    R & L & Hi & \cellcolor[HTML]{ffdaff}{\texttt{\#ffdaff}} & \cellcolor[HTML]{ffb6eb}{\texttt{\#ffb6eb}} & \cellcolor[HTML]{ff91cf}{\texttt{\#ff91cf}} & \cellcolor[HTML]{ff68b8}{\texttt{\#ff68b8}} & \cellcolor[HTML]{ff31a3}{\texttt{\#ff31a3}} & \cellcolor[HTML]{ff008e}{\texttt{\#ff008e}} & \cellcolor[HTML]{ff0078}{\texttt{\#ff0078}} &  \cellcolor[HTML]{de0065}{\textcolor{white}{\texttt{\#de0065}}} & \cellcolor[HTML]{bb0056}{\textcolor{white}{\texttt{\#bb0056}}} \\

      R & L & Lo & \cellcolor[HTML]{fffcff}{\texttt{\#fffcff}} & \cellcolor[HTML]{f9dae4}{\texttt{\#f9dae4}} & \cellcolor[HTML]{f4bbc8}{\texttt{\#f4bbc8}} & \cellcolor[HTML]{f39db1}{\texttt{\#f39db1}} & \cellcolor[HTML]{ef7f9c}{\texttt{\#ef7f9c}} & \cellcolor[HTML]{e26087}{\texttt{\#e26087}} & \cellcolor[HTML]{ca4571}{\textcolor{white}{\texttt{\#ca4571}}} & \cellcolor[HTML]{ab355e}{\textcolor{white}{\texttt{\#ab355e}}} & \cellcolor[HTML]{892c4f}{\textcolor{white}{\texttt{\#892c4f}}} \\

    R & D & Hi & \cellcolor[HTML]{eb7cb3}{\texttt{\#eb7cb3}} & \cellcolor[HTML]{da578f}{\texttt{\#da578f}} & \cellcolor[HTML]{cf2975}{\textcolor{white}{\texttt{\#cf2975}}} & \cellcolor[HTML]{c80062}{\textcolor{white}{\texttt{\#c80062}}} & \cellcolor[HTML]{be004f}{\textcolor{white}{\texttt{\#be004f}}} & \cellcolor[HTML]{ad003d}{\textcolor{white}{\texttt{\#ad003d}}} & \cellcolor[HTML]{92002b}{\textcolor{white}{\texttt{\#92002b}}} & \cellcolor[HTML]{73001c}{\textcolor{white}{\texttt{\#73001c}}} & \cellcolor[HTML]{55000a}{\textcolor{white}{\texttt{\#55000a}}} \\

    R & D & Lo & \cellcolor[HTML]{a59fad}{\texttt{\#a59fad}} & \cellcolor[HTML]{9b8089}{\texttt{\#9b8089}} & \cellcolor[HTML]{96626f}{\texttt{\#96626f}} & \cellcolor[HTML]{93455b}{\textcolor{white}{\texttt{\#93455b}}} & \cellcolor[HTML]{8d2449}{\textcolor{white}{\texttt{\#8d2449}}} & \cellcolor[HTML]{7f0037}{\textcolor{white}{\texttt{\#7f0037}}} & \cellcolor[HTML]{680024}{\textcolor{white}{\texttt{\#680024}}} & \cellcolor[HTML]{4c0014}{\textcolor{white}{\texttt{\#4c0014}}} & \cellcolor[HTML]{340000}{\textcolor{white}{\texttt{\#340000}}} \\
    
    \midrule
    
    Y & L & Hi & \cellcolor[HTML]{ffff8d}{\texttt{\#ffff8d}} & \cellcolor[HTML]{f9ed6b}{\texttt{\#f9ed6b}} & \cellcolor[HTML]{e5da48}{\texttt{\#e5da48}} & \cellcolor[HTML]{d0c81b}{\texttt{\#d0c81b}} & \cellcolor[HTML]{beb600}{\texttt{\#beb600}} & \cellcolor[HTML]{aca500}{\texttt{\#aca500}} & \cellcolor[HTML]{9a9500}{\texttt{\#9a9500}} & \cellcolor[HTML]{8a8500}{\texttt{\#8a8500}} & \cellcolor[HTML]{7b7600}{\texttt{\#7b7600}} \\
     
     Y & L & Lo & \cellcolor[HTML]{fffef3}{\texttt{\#fffef3}} & \cellcolor[HTML]{f0ebd2}{\texttt{\#f0ebd2}} & \cellcolor[HTML]{ddd8b3}{\texttt{\#ddd8b3}} & \cellcolor[HTML]{ccc695}{\texttt{\#ccc695}} & \cellcolor[HTML]{bbb379}{\texttt{\#bbb379}} & \cellcolor[HTML]{aca25e}{\texttt{\#aca25e}} & \cellcolor[HTML]{9c9245}{\texttt{\#9c9245}} & \cellcolor[HTML]{8d832c}{\texttt{\#8d832c}} & \cellcolor[HTML]{7f7413}{\texttt{\#7f7413}} \\
    
    Y & D & Hi & \cellcolor[HTML]{857c00}{\textcolor{white}{\texttt{\#857c00}}} & \cellcolor[HTML]{6f6c00}{\textcolor{white}{\texttt{\#6f6c00}}} & \cellcolor[HTML]{5c5b00}{\textcolor{white}{\texttt{\#5c5b00}}} & \cellcolor[HTML]{4c4c00}{\textcolor{white}{\texttt{\#4c4c00}}} & \cellcolor[HTML]{3f3c00}{\textcolor{white}{\texttt{\#3f3c00}}} & \cellcolor[HTML]{352e00}{\textcolor{white}{\texttt{\#352e00}}} & \cellcolor[HTML]{2f2000}{\textcolor{white}{\texttt{\#2f2000}}} & \cellcolor[HTML]{2b1400}{\textcolor{white}{\texttt{\#2b1400}}} & \cellcolor[HTML]{250100}{\textcolor{white}{\texttt{\#250100}}} \\
   
    Y & D & Lo & \cellcolor[HTML]{7f7a70}{\texttt{\#7f7a70}} & \cellcolor[HTML]{6d6953}{\textcolor{white}{\texttt{\#6d6953}}} & \cellcolor[HTML]{5d5939}{\textcolor{white}{\texttt{\#5d5939}}} & \cellcolor[HTML]{4d4920}{\textcolor{white}{\texttt{\#4d4920}}} & \cellcolor[HTML]{3e3b04}{\textcolor{white}{\texttt{\#3e3b04}}} & \cellcolor[HTML]{312d00}{\textcolor{white}{\texttt{\#312d00}}} & \cellcolor[HTML]{281f00}{\textcolor{white}{\texttt{\#281f00}}} & \cellcolor[HTML]{211300}{\textcolor{white}{\texttt{\#211300}}} & \cellcolor[HTML]{1a0000}{\textcolor{white}{\texttt{\#1a0000}}} \\
    
    \midrule
    
    G & L & Hi & \cellcolor[HTML]{83fff3}{\texttt{\#83fff3}} & \cellcolor[HTML]{15ffe2}{\texttt{\#15ffe2}} & \cellcolor[HTML]{00ead0}{\texttt{\#00ead0}} & \cellcolor[HTML]{00d8c0}{\texttt{\#00d8c0}} & \cellcolor[HTML]{00c6b1}{\texttt{\#00c6b1}} & \cellcolor[HTML]{00b19d}{\texttt{\#00b19d}} & \cellcolor[HTML]{009786}{\texttt{\#009786}} & \cellcolor[HTML]{007f6f}{\textcolor{white}{\texttt{\#007f6f}}} & \cellcolor[HTML]{006856}{\textcolor{white}{\texttt{\#006856}}} \\
    G & L & Lo & \cellcolor[HTML]{f7fff3}{\texttt{\#f7fff3}} & \cellcolor[HTML]{c7eae2}{\texttt{\#c7eae2}} & \cellcolor[HTML]{93d7d0}{\texttt{\#93d7d0}} & \cellcolor[HTML]{54c8c0}{\texttt{\#54c8c0}} & \cellcolor[HTML]{00b9b0}{\texttt{\#00b9b0}} & \cellcolor[HTML]{00a59c}{\texttt{\#00a59c}} & \cellcolor[HTML]{008e85}{\textcolor{white}{\texttt{\#008e85}}} & \cellcolor[HTML]{00766e}{\textcolor{white}{\texttt{\#00766e}}} & \cellcolor[HTML]{005f55}{\textcolor{white}{\texttt{\#005f55}}} \\
    
    G & D & Hi & \cellcolor[HTML]{00b896}{\texttt{\#00b896}} & \cellcolor[HTML]{009f87}{\texttt{\#009f87}} & \cellcolor[HTML]{008a77}{\textcolor{white}{\texttt{\#008a77}}} & \cellcolor[HTML]{007969}{\textcolor{white}{\texttt{\#007969}}} & \cellcolor[HTML]{00675b}{\textcolor{white}{\texttt{\#00675b}}} & \cellcolor[HTML]{00524a}{\textcolor{white}{\texttt{\#00524a}}} & \cellcolor[HTML]{003f36}{\textcolor{white}{\texttt{\#003f36}}} & \cellcolor[HTML]{003221}{\textcolor{white}{\texttt{\#003221}}} & \cellcolor[HTML]{002809}{\textcolor{white}{\texttt{\#002809}}} \\
    
    G & D & Lo & \cellcolor[HTML]{9aa496}{\texttt{\#9aa496}} & \cellcolor[HTML]{6d8e87}{\texttt{\#6d8e87}} & \cellcolor[HTML]{397c77}{\textcolor{white}{\texttt{\#397c77}}} & \cellcolor[HTML]{006e68}{\textcolor{white}{\texttt{\#006e68}}} & \cellcolor[HTML]{005f5a}{\textcolor{white}{\texttt{\#005f5a}}} & \cellcolor[HTML]{004c49}{\textcolor{white}{\texttt{\#004c49}}} & \cellcolor[HTML]{003835}{\textcolor{white}{\texttt{\#003835}}} & \cellcolor[HTML]{002821}{\textcolor{white}{\texttt{\#002821}}} & \cellcolor[HTML]{001a09}{\textcolor{white}{\texttt{\#001a09}}} \\
    
    \midrule
    
    B & L & Hi & \cellcolor[HTML]{c7ffff}{\texttt{\#c7ffff}} & \cellcolor[HTML]{97e8ff}{\texttt{\#97e8ff}} & \cellcolor[HTML]{46d3ff}{\texttt{\#46d3ff}} & \cellcolor[HTML]{00c3ff}{\texttt{\#00c3ff}} & \cellcolor[HTML]{00b3ff}{\texttt{\#00b3ff}} & \cellcolor[HTML]{00a1ff}{\texttt{\#00a1ff}} & \cellcolor[HTML]{008aff}{\texttt{\#008aff}} & \cellcolor[HTML]{0074fe}{\texttt{\#0074fe}} & \cellcolor[HTML]{005ed8}{\textcolor{white}{\texttt{\#005ed8}}} \\
    B & L & Lo & \cellcolor[HTML]{feffff}{\texttt{\#feffff}} & \cellcolor[HTML]{dae3f5}{\texttt{\#dae3f5}} & \cellcolor[HTML]{afceef}{\texttt{\#afceef}} & \cellcolor[HTML]{7dbeee}{\texttt{\#7dbeee}} & \cellcolor[HTML]{3baeea}{\texttt{\#3baeea}} & \cellcolor[HTML]{009bdf}{\texttt{\#009bdf}} & \cellcolor[HTML]{0085c8}{\textcolor{white}{\texttt{\#0085c8}}} & \cellcolor[HTML]{006ea9}{\textcolor{white}{\texttt{\#006ea9}}} & \cellcolor[HTML]{005987}{\textcolor{white}{\texttt{\#005987}}} \\
    B & D & Hi & \cellcolor[HTML]{62a6f9}{\texttt{\#62a6f9}} & \cellcolor[HTML]{108cee}{\texttt{\#108cee}} & \cellcolor[HTML]{007ae6}{\texttt{\#007ae6}} & \cellcolor[HTML]{006ce4}{\texttt{\#006ce4}} & \cellcolor[HTML]{005ddf}{\textcolor{white}{\texttt{\#005ddf}}} & \cellcolor[HTML]{004dd4}{\textcolor{white}{\texttt{\#004dd4}}} & \cellcolor[HTML]{003abc}{\textcolor{white}{\texttt{\#003abc}}} & \cellcolor[HTML]{00299d}{\textcolor{white}{\texttt{\#00299d}}} & \cellcolor[HTML]{00197b}{\textcolor{white}{\texttt{\#00197b}}} \\
    B & D & Lo & \cellcolor[HTML]{a0a1a2}{\texttt{\#a0a1a2}} & \cellcolor[HTML]{7f8898}{\texttt{\#7f8898}} & \cellcolor[HTML]{567592}{\textcolor{white}{\texttt{\#567592}}} & \cellcolor[HTML]{156691}{\textcolor{white}{\texttt{\#156691}}} & \cellcolor[HTML]{00588d}{\textcolor{white}{\texttt{\#00588d}}} & \cellcolor[HTML]{004783}{\textcolor{white}{\texttt{\#004783}}} & \cellcolor[HTML]{00336e}{\textcolor{white}{\texttt{\#00336e}}} & \cellcolor[HTML]{002253}{\textcolor{white}{\texttt{\#002253}}} & \cellcolor[HTML]{001035}{\textcolor{white}{\texttt{\#001035}}} \\
    \bottomrule
    \end{tabular}
    \label{tab:color_scales_values}
\end{table*}

\begin{figure*}[t!]
    \centering
    \includegraphics[width=1\linewidth]{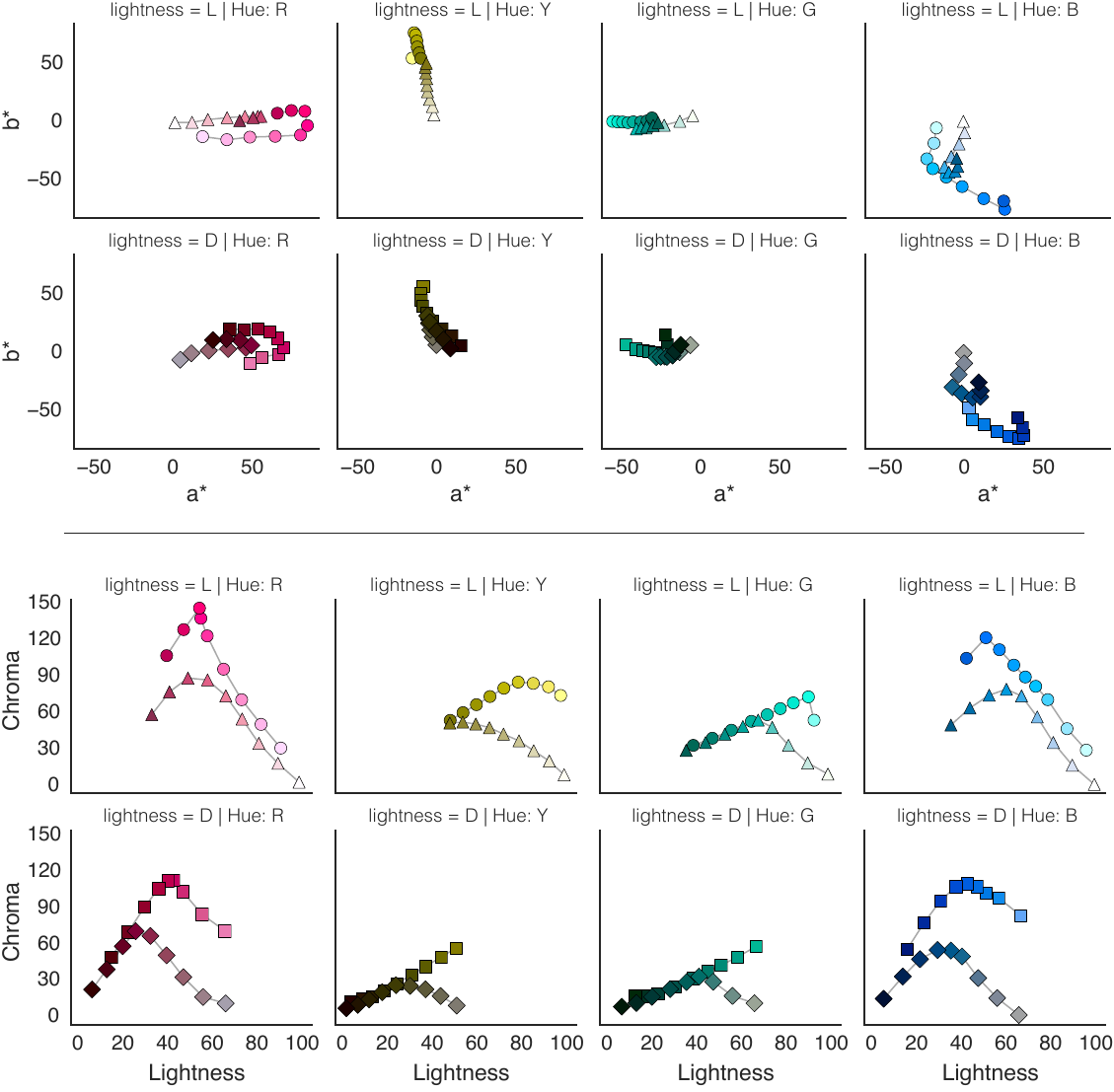}
    \caption{Each of the 16 color scales plotted according to their a* (roughly red/green) vs. b* (roughly yellow/blue) coordinates in CIELAB space (top) and according to their L* (lightness) and C* (chroma) coordinates in CIELCh space (bottom). Note: `L*' is the same in CIELAB and CIELCh spaces. 
    Each facet depicts both the high chroma and low chroma color scales within a given lightness level with colors belonging to the same color scale being connected by a line. The coordinates shown here appear different from the coordinates plotted within the Color Crafter tool, with discrepancies likely due to clipping and rounding issues not visualized in the tool. Marker shapes use the same mappings used in Exp. 1 (Fig. \ref{fig:exp1_assoc}) and Exp. 2 (Fig. \ref{fig:Exp2_assoc}). Circles: light+high chroma, triangles: light+low chroma, squares: dark+high chroma, diamonds: dark+low chroma.}
    \label{fig:color-scale-coordinates}
\end{figure*}

\begin{figure*}[t!]
    \centering
    \includegraphics[width=1\linewidth]{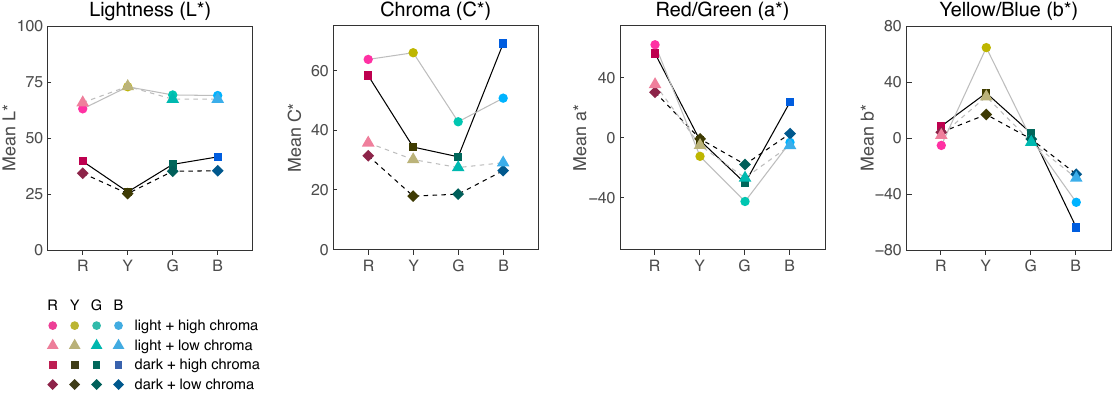}
    \caption{The mean L*, C*, a*, and b* coordinates for each color scale as used in the analyses of the association data, plotted in the same way as the association data shown in Figs. \ref{fig:exp1_assoc} and \ref{fig:Exp2_assoc}. The values correspond to the coordinates in Table \ref{tab:ColorScaleCoords}.}

    \label{fig:color_scale_means}
\end{figure*}

\begin{figure*}[ht!]
 \centering
 \includegraphics[width=\textwidth]{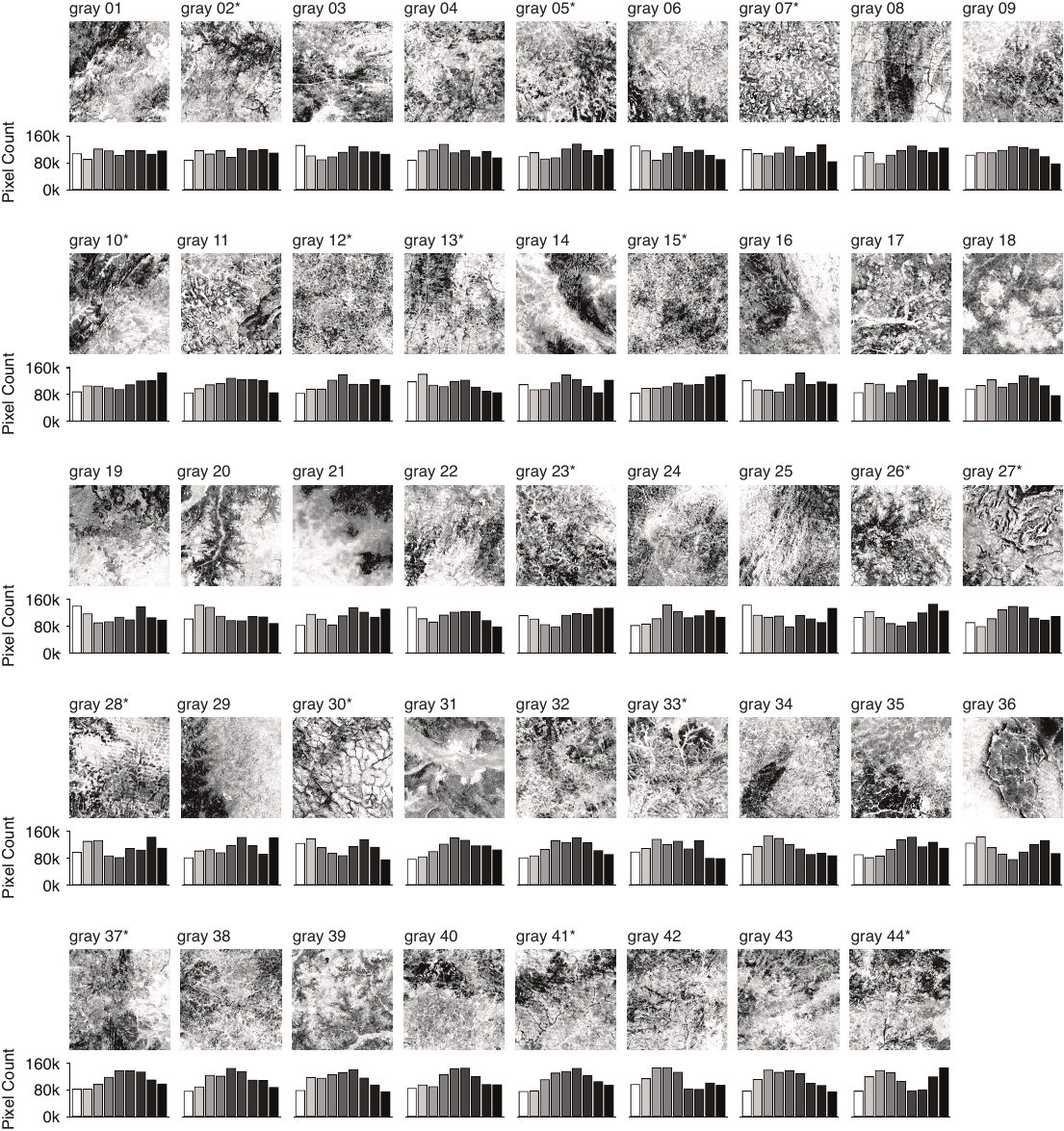}
 \caption{Grayscale colormaps of the 44 subtiles selected for the Exp. 1 pilot. The colormaps show biomass density values scaled to range from 0 to 100, grouped into 9 bins, and then mapped to a 9-step gray color scale (see main text for details). The histogram below each colormap shows the number of pixels in each of the 9 bins, with each bar color in the histogram corresponding to the pixel colors assigned to values in that bar's bin. The maps are labeled 01 to 44 in order from most to least uniform distribution within the set. An asterisk next to a map number indicates that map was selected for Exp. 1.}
 \label{fig:graymaps44}
 \vspace{-6mm}
\end{figure*}

\begin{figure*}[ht!]
% \centering
 \includegraphics[width=1.0\textwidth]{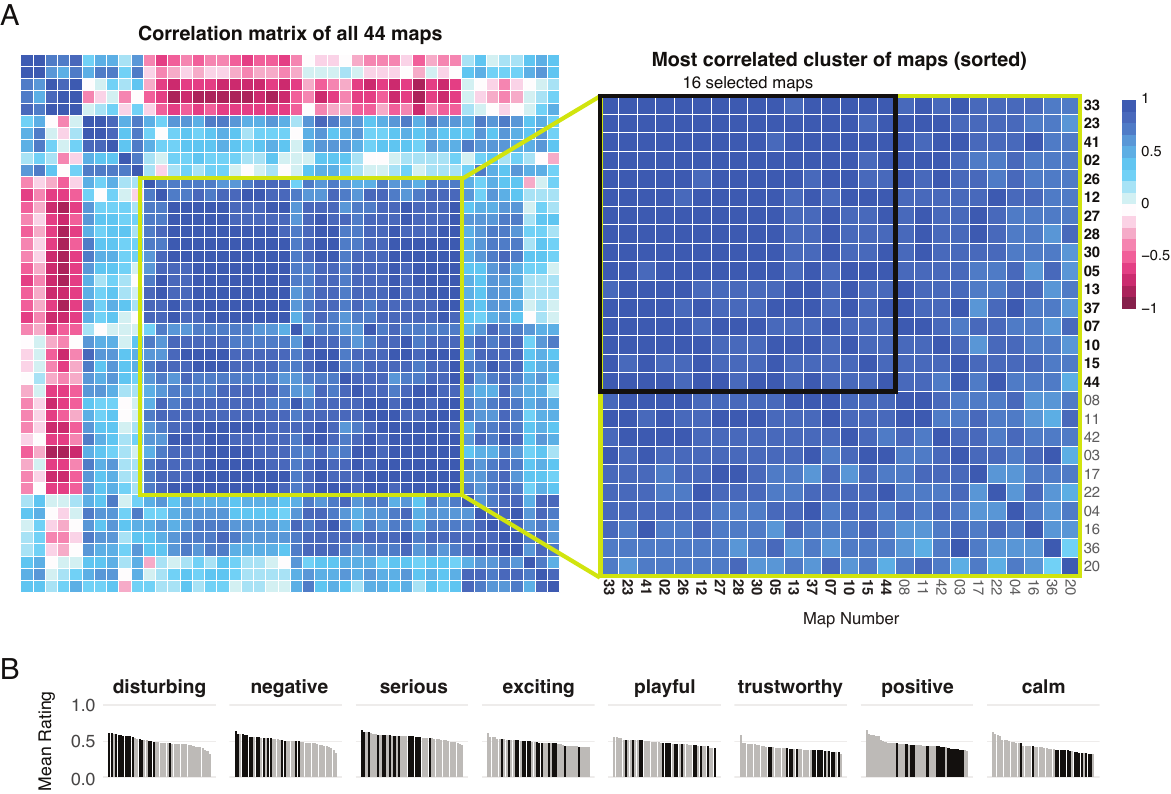}
 \caption{(A) Pairwise correlations between all 44 grayscale maps in terms of their mean associations with each of the 8 emotion concepts. The inset highlights the subset of 26 highly correlated maps with the black border indicating the final set of 16 selected maps. (B) Mean association ratings between each emotion concept and the 44 grayscale maps sorted from high to low. Black bars correspond to the 16 selected maps.}
 \label{fig:gray_emo}
\end{figure*}

\begin{table*}[ht!]
    \centering
     \caption{The mean a*, b*, C*, and L* values for each of the 16 color scales.}
    \begin{tabular}{lll|cccc}
        \toprule
        \textbf{Hue} & \textbf{Lightness} & \textbf{Chroma} & \textbf{a*} & \textbf{b*} & \textbf{C*} & \textbf{L*} \\         \midrule
        R & L & Hi & 1.93434247 & -0.1454748361 & 63.72081 & 63.10979 \\
        R & L & Lo & 1.04967609 & 0.0901855282 & 35.76330 & 65.88214 \\
        R & D & Hi & 1.73771836 & 0.3027289812 & 58.28588 & 39.65978 \\
        R & D & Lo & 0.87749921 & 0.1607637458 & 31.44501 & 34.33223 \\
        \midrule
        Y & L & Hi & -0.54932776 & 2.1278762379 & 65.95778 & 72.95918 \\
        Y & L & Lo & -0.30369870 & 0.9886564320 & 30.15460 & 73.13807 \\ 
        Y & D & Hi & -0.19653845 & 1.0734107631 & 34.35054 & 25.98140 \\       
        Y & D & Lo & -0.16100056 & 0.5724112557 & 17.91222 & 25.22744 \\
        
        \midrule
        
        G & L & Hi & -1.55429718 & 0.0032070844 & 42.82724 & 69.30993 \\
        G & L & Lo & -1.03120908 & -0.0694310119 & 27.49006 & 67.42741 \\
        G & D & Hi & -1.14410813 & 0.1328061352 & 31.09054 & 38.28996 \\
        G & D & Lo & -0.73006326 & 0.0009802904 & 18.57621 & 35.20695 \\
        
        \midrule
        B & L & High & -0.23798045 & -1.4679156558 & 50.74219 & 69.05558 \\
        B & L & Low & -0.30639061 & -0.9043147546 & 29.10485 & 67.42697 \\
        B & D & Low & -0.04389614 & -0.8139315872 & 26.44173 & 35.41818 \\
        B & D & High & 0.65927419 & -2.0519586083 & 69.03252 & 41.64495 \\
        
        \bottomrule
        
    \end{tabular}
    \label{tab:ColorScaleCoords}
\end{table*}

\begin{table*}[ht]
    \centering
     \caption{The mean a*, b*, C*, and L* values across colormaps constructed from each of 16 color scales applied to the light or dark shifted underlying datasets. }
    \begin{tabular}{llll|cccc}
        \toprule
        \textbf{Hue} & \textbf{Lightness} & \textbf{Chroma} & \textbf{Shift}& \textbf{a*} & \textbf{b*} & \textbf{C*} & \textbf{L*} \\         \midrule

         R & L & Hi & lightshift & 40.0135614 & -10.6921325 & 42.87100 & 78.20307 \\
        R & L & Hi & darkshift & 67.5073924 & 2.6699313 & 68.19989 & 48.63701 \\
       
        R & L & Lo & lightshift & 14.7269739 & 0.1099563 & 14.96044 & 86.71514 \\
        R & L & Lo & darkshift & 42.4642460 & 1.4143864 & 42.51972 & 43.47099 \\
        
        R & D & Hi & lightshift & 54.0810480 & -2.0704807 & 55.19350 & 54.97053 \\
        R & D & Hi & darkshift & 45.3967416 & 16.3722232 & 49.18207 & 24.84947 \\
        
        R & D & Lo & lightshift & 14.7977216 & -2.4443576 & 16.81123 & 55.12965 \\
        R & D & Lo & darkshift & 29.7242735 & 8.0089109 & 31.25016 & 18.29705 \\
        
        \midrule

        Y & L & Hi & lightshift & -14.0808327 & 59.7983622 & 61.50356 & 88.30894 \\
        Y & L & Hi & darkshift & -10.7908360 & 57.9009221 & 58.90664 & 56.06223 \\
        
        Y & L & Lo & lightshift & -2.7583008 & 13.2590314 & 13.56206 & 91.11995 \\
        Y & L & Lo & darkshift & -6.0373247 & 41.8980402 & 42.34626 & 58.31579 \\
       
        Y & D & Hi & lightshift & -6.1480114 & 45.7094150 & 46.84775 & 39.77389 \\
        Y & D & Hi & darkshift & 9.6213069 & 14.4209140 & 21.60028 & 10.63624 \\
    
        Y & D & Lo & lightshift & -0.9956447 & 12.3967659 & 12.81686 & 39.63490 \\
        Y & D & Lo & darkshift & 5.0790834 & 8.8927506 & 12.99972 & 10.06670 \\

        \midrule

        G & L & Hi & lightshift & -41.6368678 & -3.1385800 & 41.84410 & 86.28464 \\ 
        G & L & Hi & darkshift & -35.3379998 & 1.2924016 & 35.42387 & 50.46052 \\
        
        G & L & Lo & lightshift & -14.4266199 & 1.1910087 & 15.63487 & 87.17643 \\
        G & L & Lo & darkshift & -27.6476810 & -1.9855748 & 27.82384 & 45.28313 \\

        G & D & Hi & lightshift & -40.7530573 & 4.5757776 & 41.16170 & 56.36026 \\
        G & D & Hi & darkshift & -23.8534733 & 10.4018087 & 26.83526 & 19.70521 \\
        
        G & D & Lo & lightshift & -11.6968128 & 2.8149495 & 13.28014 & 53.27698 \\
        G & D & Lo & darkshift & -15.3398031 & 2.7360430 & 16.33281 & 16.31158 \\

        \midrule
        B & L & Hi & lightshift & -13.5420398 & -22.0317963 & 31.21908 & 85.40398 \\
        B & L & Hi & darkshift & 15.9520056 & -62.5756042 & 66.78684 & 51.98345 \\

        B & L & Lo & lightshift & -2.1692914 & -11.8800853 & 12.23605 & 87.02787 \\
        B & L & Lo & darkshift & -5.0518189 & -32.3520955 & 32.79897 & 46.64972 \\
       
        B & D & Hi & lightshift & 10.4725279 & -54.9298526 & 56.67350 & 57.75702 \\
        B & D & Hi & darkshift & 31.1647432 & -60.3134036 & 68.42741 & 26.22301 \\
        
        B & D & Lo & lightshift & 0.7338501 & -12.6481391 & 12.96321 & 51.99386 \\
        B & D & Lo & darkshift & 7.5448208 & -27.5347422 & 28.89766 & 15.63829 \\

        \bottomrule
        
    \end{tabular}
    \label{tab:my_label}
\end{table*}

\begin{table*}[ht!]
\centering
\caption{Experiment 1 mixed effects model results. Predictors were $z$-scored. (Stars are omitted from L*, C*, a*, and b* for table readability.)}
\begin{tabular}{lrrrrr|lrrrrr}
\midrule
& & \textbf{Positive} & & & & & & \textbf{Exciting} & & \\
\midrule
\textbf{Predictor} & \boldsymbol{$\beta$} & \boldsymbol{$df$} & \textbf{SE} & \boldsymbol{$t$} & \boldsymbol{$p$} & \textbf{Predictor} & \boldsymbol{$\beta$} & \boldsymbol{$df$} & \textbf{SE} & \boldsymbol{$t$} & \boldsymbol{$p$}\\
\midrule
(Intercept) & 0.532 & 40.9  & 0.021 & 25.87 & 0.0000 &   (Intercept) & 0.530 & 41.2  & 0.021 & 25.30 & 0.0000 \\
L & 0.116 & 39.0  & 0.021 & 5.47 & 0.0000 &              L & 0.120 & 42.5  & 0.019 & 6.34 & 0.0000 \\
C & 0.059 & 104.0  & 0.012 & 5.08 & 0.0000 &              C & 0.070 & 72.1  & 0.014 & 4.85 & 0.0000 \\
a & -0.029 & 43.6  & 0.014 & -2.06 & 0.0458 &            a & -0.014 & 64.7 & 0.010 & -1.37 & 0.1745 \\
b & -0.067 & 44.4  & 0.014 & -4.83 & 0.0000 &            b & -0.080 & 47.6  & 0.013 & -6.07 & 0.0000 \\
L:C & -0.019 & 861.0 & 0.009 & -2.07 & 0.0386 &          L:C & -0.012 & 857.0 & 0.009 & -1.29 & 0.1958 \\
L:a & -0.017 & 861.0 & 0.009 & -2.00 & 0.0456 &          L:a & -0.019 & 857.0 & 0.009 & -2.16 & 0.0310 \\
L:b & -0.015 & 861.0 & 0.007 & -2.15 & 0.0317 &          L:b & -0.038 & 857.0 & 0.007 & -5.32 & 0.0000 \\

C:a & 0.034 & 861.0 & 0.008 & 4.19 & 0.0000 &            C:a & 0.043 & 857.0  & 0.008 & 5.12 & 0.0000 \\
C:b & 0.021 & 861.0 & 0.006 & 3.23 & 0.0013 &            C:b & 0.033 & 857.0 & 0.007 & 5.03 & 0.0000 \\
L:C:a & 0.008 & 861.0 & 0.009 & 0.83 & 0.4086 &          L:C:a & -0.000 & 857.0 & 0.010 & -0.05 & 0.9622 \\
L:C:b & -0.002 & 861.0 & 0.007 & -0.35 & 0.7248 &        L:C:b & -0.009 & 857.0 & 0.007 & -1.30 & 0.1936 \\
\toprule
& & \textbf{Playful} & & & & & & \textbf{Negative} & & \\
\midrule
\textbf{Predictor} & \boldsymbol{$\beta$} & \boldsymbol{$df$} & \textbf{SE} & \boldsymbol{$t$} & \boldsymbol{$p$} & \textbf{Predictor} & \boldsymbol{$\beta$} & \boldsymbol{$df$} & \textbf{SE} & \boldsymbol{$t$} & \boldsymbol{$p$}\\
\midrule
(Intercept) & 0.511 & 48.0  & 0.018 & 27.64 & 0.0000 & (Intercept) & 0.438 & 39.4  & 0.023 & 19.12 & 0.0000 \\
L & 0.132 & 40.5  & 0.022 & 6.06 & 0.0000 & L & -0.109 & 43.7   & 0.018 & -5.99 & 0.0000 \\
C & 0.044 & 78.1  & 0.015 & 2.99 & 0.0037 & C & -0.049 & 71.1   & 0.014 & -3.49 & 0.0008 \\

a & -0.003 & 49.9  & 0.013 & -0.22 & 0.8237 & a & 0.026 & 40.3   & 0.016 & 1.68 & 0.1009 \\
b & -0.084 & 45.2  & 0.015 & -5.65 & 0.0000 & b & 0.060 & 42.9   & 0.015 & 4.00 & 0.0002 \\

L:C & -0.008 & 857.0 & 0.010 & -0.78 & 0.4342 & L:C & 0.027 & 853.0 & 0.009 & 2.88 & 0.0041 \\
L:a & -0.022 & 857.0 & 0.010 & -2.25 & 0.0247 & L:a & -0.015 & 853.0 & 0.009 & -1.64 & 0.1017 \\
L:b & -0.017 & 857.0 & 0.008 & -2.21 & 0.0272 & L:b & 0.016 & 853.0 & 0.007 & 2.29 & 0.0223 \\
C:a & 0.039 & 857.0 & 0.009 & 4.35 & 0.0000 & C:a & -0.022 & 853.0 & 0.008 & -2.55 & 0.0110 \\
C:b & 0.012 & 857.0 & 0.007 & 1.74 & 0.0817 & C:b & -0.019 & 853.0 & 0.007 & -2.78 & 0.0056 \\
L:C:a & 0.006 &857.0 & 0.010 & 0.60 & 0.5487 & L:C:a & 0.010 & 853.0 & 0.010 & 0.98 & 0.3284 \\
L:C:b & -0.005 & 857.0 & 0.008 & -0.70 & 0.4856 & L:C:b & 0.008 & 853.0 & 0.007 & 1.04 & 0.2973 \\

\toprule
& & \textbf{Disturbing} & & & & & & \textbf{Serious} & & \\
\midrule
\textbf{Predictor} & \boldsymbol{$\beta$} & \boldsymbol{$df$} & \textbf{SE} & \boldsymbol{$t$} & \boldsymbol{$p$} & \textbf{Predictor} & \boldsymbol{$\beta$} & \boldsymbol{$df$} & \textbf{SE} & \boldsymbol{$t$} & \boldsymbol{$p$}\\
\midrule
(Intercept) & 0.438 & 40.0  & 0.023 & 18.82 & 0.0000 & (Intercept) & 0.527 & 38.6  & 0.024 & 22.11 & 0.0000 \\
L & -0.062 & 48.3  & 0.017 & -3.64 & 0.0007 & L & -0.144 & 40.4  & 0.021 & -6.95 & 0.0000 \\
C & -0.039 & 79.1  & 0.014 & -2.74 & 0.0076 & C & -0.036 & 103.0  & 0.012 & -2.88 & 0.0048 \\
a & 0.044 & 46.0  & 0.014 & 3.27 & 0.0020 & a & -0.013 & 55.6 & 0.011 & -1.17 & 0.2473 \\
b & 0.098 & 40.5  & 0.017 & 5.71 & 0.0000 & b & 0.029 & 44.0  & 0.015 & 1.97 & 0.0553 \\
L:C & 0.024 & 848.0 & 0.010 & 2.39 & 0.0172 & L:C & -0.001 & 860.0& 0.009 & -0.08 & 0.9358 \\
L:a & -0.004 & 848.0 & 0.010 & -0.40 & 0.6904 & L:a & 0.011 & 860.0 & 0.009 & 1.20 & 0.2319 \\
L:b & 0.011 & 848.0 & 0.007 & 1.46 & 0.1437 & L:b & 0.030 & 860.0 & 0.007 & 4.19 & 0.0000 \\
C:a & -0.036 & 848.0 & 0.009 & -4.11 & 0.0000 & C:a & -0.026 & 860.0 & 0.008 & -3.13 & 0.0018 \\
C:b & -0.035 & 848.0 & 0.007 & -5.05 & 0.0000 & C:b & -0.017 & 860.0 & 0.007 & -2.55 & 0.0111 \\
L:C:a & 0.003 &848.0 & 0.010 & 0.33 & 0.7389 & L:C:a & 0.002 & 860.0 & 0.010 & 0.16 & 0.8768 \\
L:C:b & 0.004 & 848.0 & 0.008 & 0.51 & 0.6104 & L:C:b & 0.002 & 860.0 & 0.007 & 0.22 & 0.8259 \\
\toprule
& & \textbf{Calm} & & & & & & \textbf{Trustworthy} & & \\
\midrule
\textbf{Predictor} & \boldsymbol{$\beta$} & \boldsymbol{$df$} & \textbf{SE} & \boldsymbol{$t$} & \boldsymbol{$p$} & \textbf{Predictor} & \boldsymbol{$\beta$} & \boldsymbol{$df$} & \textbf{SE} & \boldsymbol{$t$} & \boldsymbol{$p$}\\
\midrule
(Intercept) & 0.513 & 49.4  & 0.019 & 27.55 & 0.0000 & (Intercept) & 0.509 & 40.9  & 0.021 & 24.51 & 0.0000 \\

L & 0.035 & 38.9  & 0.025 & 1.42 & 0.1627 & L & 0.075 & 42.3  & 0.019 & 4.02 & 0.0002 \\
C & -0.015 & 69.8  & 0.016 & -0.95 & 0.3435 & C & 0.012 & 79.0  & 0.013 & 0.90 & 0.3714 \\
a & -0.036 & 49.1  & 0.013 & -2.73 & 0.0088 & a & -0.014 & 48.7  & 0.012 & -1.20 & 0.2366 \\
b & -0.104 & 45.2  & 0.016 & -6.70 & 0.0000 & b & -0.091 & 41.7  & 0.015 & -5.97 & 0.0000 \\

L:C & -0.013 & 846.0 & 0.010 & -1.22 & 0.2224 & L:C & -0.020 & 854.0 & 0.009 & -2.15 & 0.0321 \\

L:a & 0.005 & 846.0 & 0.010 & 0.53 & 0.5944 & L:a & -0.002 & 854.0 & 0.009 & -0.27 & 0.7842 \\
L:b & 0.011 & 846.0& 0.008 & 1.45 & 0.1484 & L:b & -0.005 & 854.0 & 0.007 & -0.73 & 0.4635 \\

C:a & -0.002 & 846.0 & 0.009 & -0.19 & 0.8479 & C:a & 0.014 & 854.0 & 0.008 & 1.74 & 0.0829 \\
C:b& 0.005 & 846.0 & 0.007 & 0.70 & 0.4828 & C:b & 0.014 & 854.0 & 0.007 & 2.14 & 0.0327 \\
L:C:a & -0.024& 846.0 & 0.011 & -2.23 & 0.0263 & L:C:a & -0.008 & 854.0 & 0.009 & -0.87 & 0.3839 \\
L:C:b & 0.014 & 846.0 & 0.008 & 1.70 & 0.0895 & L:C:b & 0.004 & 854.0 & 0.007 & 0.61 & 0.5401 \\
\hline
\end{tabular}
\label{tab:exp1-lmer-results-2col}
\end{table*}

 \begin{figure*}[ht!]
  \centering
 \includegraphics[width=0.9\textwidth]{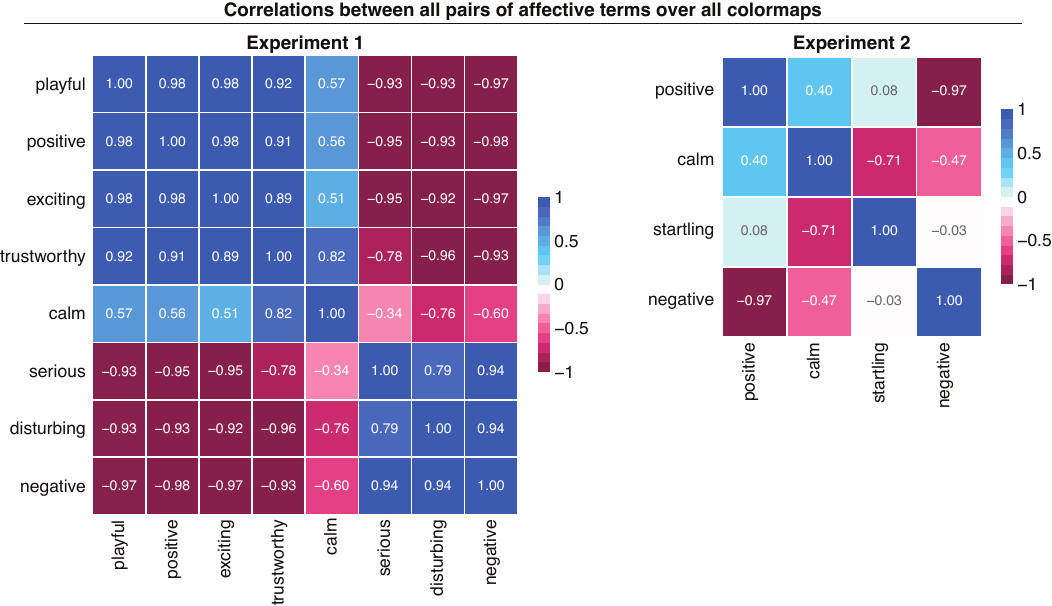}
  \caption{Correlation for each pairwise combination of affective terms across all colormaps in Experiment 1 (left) and Experiment 2 (right). }
  \label{fig:concept-concept-corr}
  \vspace{-0mm}
 \end{figure*}

{
\begin{table*}[ht!]
% \captionsetup{skip=10pt} 
    \centering
\caption{Results from multiple linear regression models predicting emotion space principal component scores from color space coordinates (see main text for details). Predictors were $z$-scored.  }
\begin{tabular}{lll|ccccc}
\toprule

\textbf{Experiment} & \textbf{PC} & \textbf{Predictor} & \boldsymbol{$\beta$} &  \textbf{SE} &  \boldsymbol{$t$} &  \boldsymbol{$p$} \\
\midrule
Exp. 1 &PC1 & Intercept& 0.0  & 0.182 & 0.00 & 1.000 \\
 && L* & 1.902  & 0.211 & 8.99 & <0.001 \\
 && C* & 0.573 & 0.235 & 2.44 & 0.033 \\
 && a* & -0.143  & 0.225 & -0.635 & 0.538 \\
 && b* & -1.407 & 0.190 & -7.398 & <0.001 \\

 \midrule
&PC2 & Intercept& 0.0& 0.109 & 0.00 & 1.000 \\
&& L* & 0.3471 &  0.126 & 3.092 & 0.028 \\
&& C* & 0.087 &  0.140 & 2.203 & 0.545 \\
&& a* & 0.506 & 0.134 & 3.77 & 0.003 \\
&& b* & 0.649 & 0.113 & 4.50 & <0.001 \\

\toprule

Exp.2 & PC1 & Intercept& 0.0  & 0.144 & 0.00 & 1.000 \\
&&L* & 0.697 &  0.149 & 4.659 & <0.001 \\
&&C* & 0.6026 & 0.157 & 2.988 & <0.001 \\
&& a* & -0.386 &  0.163 & -2.373 & 0.025 \\
&& b* & -0.966 &  0.149 & -6.483 & <0.001 \\

 \midrule
 
& PC2 & Intercept& 0.0 & 0.1269 & 0.00 & 1.000 \\
&&L* & 0.709 & 0.1269 & 5.589 & <0.001 \\
&&C* & 0.208 &  0.1269 & 1.641 & 0.053 \\
&& a* & 0.698 &  0.1794 & 3.888 & <0.001 \\
&& b* & 0.673 & 0.1794 & 3.751 & <0.001 \\

\bottomrule
\end{tabular}

 \label{tab:exp1-pc-regression}
\end{table*}
}

\begin{figure*}[ht!!]
 \centering
 \includegraphics[width=0.9\textwidth]{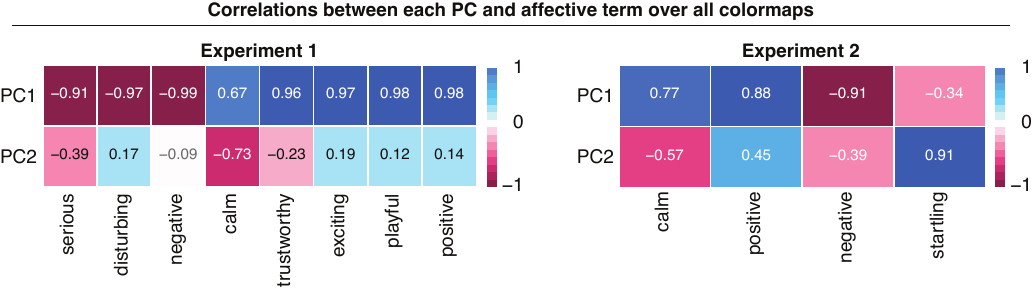}
 \caption{Correlations between mean affective concept ratings and PC loadings across all colormaps in Experiment 1 (left) and Experiment 2 (right). }
 \label{fig:emotion-PC-corr}
 % \vspace{-0mm}
\end{figure*}

\begin{figure*}[ht!!]
 \centering
 \includegraphics[width=0.75\textwidth]{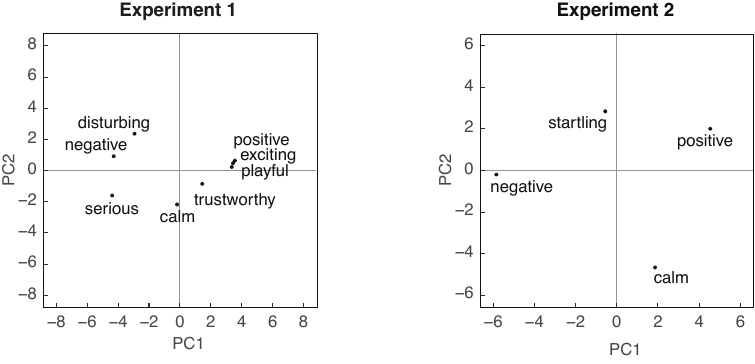}
 \caption{Organization of emotion concepts along the first two principal components derived from the mean \textbf{color scale} $\times$ \textbf{emotion concept} matrices in Experiment 1 (left) and Experiment 2 (right).}
 \label{fig:emotion-concept-PCA}
 % \vspace{-0mm}
\end{figure*}

\begin{table*}[ht!]
\centering
\caption{Experiment 2 mixed effects model results. Predictors were $z$-scored. (Stars are omitted from L*, C*, a*, and b* for table readability.)}
\begin{tabular}{lccccc|lcccccc}
\midrule
& & \textbf{Positive} & & & & & & \textbf{Negative} & & \\
\midrule
\textbf{Predictor} & \boldsymbol{$\beta$} & \boldsymbol{$df$} & \textbf{SE} & \boldsymbol{$t$} & \boldsymbol{$p$} & \textbf{Predictor} & \boldsymbol{$\beta$} & \boldsymbol{$df$} & \textbf{SE} & \boldsymbol{$t$} & \boldsymbol{$p$}\\
\midrule
(Intercept) & 0.516 & 51.9  & 0.013 & 39.38 & 0.0000 & (Intercept) & 0.475 & 42.9  & 0.017 & 28.48 & 0.0000 \\
shift & 0.046 & 1990.0 & 0.007 & 6.32 & 0.0000 & shift & -0.050 & 1990.0 & 0.007 & -6.95 & 0.0000 \\
L & 0.102 & 46.3  & 0.014 & 7.11 & 0.0000 & L & -0.102 & 47.0  & 0.014 & -7.53 & 0.0000 \\
C & 0.059 & 11.4  & 0.009 & 6.34 & 0.0000 & C & -0.054 & 86.9  & 0.010 & -5.33 & 0.0000 \\
a & -0.007 & 47.6 & 0.011 & -0.63 & 0.5256 & a & 0.036 & 41.3 & 0.013 & 2.75 & 0.0059 \\
b & -0.064 & 40.8  & 0.014 & -4.53 & 0.0000 & b & 0.080 & 43.7  & 0.012 & 6.51 & 0.0000 \\

L:C & -0.022 & 1990.0 & 0.007 & -3.01 & 0.0026 & L:C & 0.022 & 1990.0 & 0.007 & 3.09 & 0.0020 \\

L:a & -0.004 & 1990.0 & 0.007 & -0.60 & 0.5469 & L:a & -0.009 & 1990.0 & 0.007 & -1.27 & 0.2038 \\
L:b & 0.003 & 1990.0 & 0.006 & 0.52 & 0.6041 & L:b & -0.006 & 1990.0 & 0.005 & -1.06 & 0.2909 \\
C:a & 0.034 & 1990.0 & 0.006 & 5.29 & 0.0000 & C:a & -0.044 & 1990.0 & 0.006 & -6.85 & 0.0000 \\
C:b & 0.032 & 1990.0 & 0.005 & 6.18 & 0.0000 & C:b & -0.039 & 1990.0 & 0.005 & -7.63 & 0.0000 \\
L:C:a & 0.010 & 1990.0 & 0.007 & 1.34 & 0.1796 & L:C:a & -0.006 & 1990.0 & 0.007 & -0.79 & 0.4312 \\
L:C:b & -0.018 & 1990.0 & 0.006 & -3.32 & 0.0009 & L:C:b & 0.011 & 1990.0 & 0.005 & 2.02 & 0.0437 \\
L:shift & -0.037 & 1990.0 & 0.007 & -5.14 & 0.0000 & L:shift & 0.024 & 1990.0 & 0.007 & 3.40 & 0.0007 \\
C:shift & 0.014 & 1990.0 & 0.008 & 1.77 & 0.0771 & C:shift & -0.002 & 1990.0 & 0.008 & -0.29 & 0.7753 \\
a:shift & -0.006 & 1990.0 & 0.006 & -0.88 & 0.3812 & a:shift & -0.005 & 1990.0 & 0.006 & -0.83 & 0.4049 \\
b:shift & 0.014 & 1990.0 & 0.006 & 2.29 & 0.0218 & b:shift & -0.019 & 1990.0 & 0.006 & -3.04 & 0.0024 \\

L:C:shift & -0.009 & 1990.0& 0.007 & -1.19 & 0.2328 & L:C:shift & 0.010 & 1990.0 & 0.007 & 1.34 & 0.1804 \\

L:a:shift & -0.000 & 1990.0 & 0.007 & -0.05 & 0.9567 & L:a:shift & 0.002 & 1990.0 & 0.007 & 0.34 & 0.7309 \\
L:b:shift & 0.006 & 1990.0 & 0.006 & 1.13 & 0.2584 & L:b:shift & -0.008 & 1990.0 & 0.005 & -1.56 & 0.1197 \\

C:a:shift & -0.001 & 1990.0 & 0.006 & -0.08 & 0.9345 & C:a:shift & -0.010 & 1990.0 & 0.006 & -1.61 & 0.1069 \\
C:b:shift & 0.003 &1990.0 & 0.005 & 0.60 & 0.5517 & C:b:shift & -0.005 & 1990.0 & 0.005 & -0.98 & 0.3261 \\
L:C:a:shift & -0.016 & 1990.0 & 0.007 & -2.19 & 0.0284 & L:C:a:shift & 0.019 & 1990.0 & 0.007 & 2.55 & 0.0109 \\
L:C:b:shift & -0.012 & 1990.0 & 0.006 & -2.19 & 0.0287 & L:C:b:shift & 0.015 & 1990.0 & 0.005 & 2.70 & 0.0070 \\

\toprule
& & \textbf{Calm} & & & & & & \textbf{Startling} & & \\
\midrule
\textbf{Predictor} & \boldsymbol{$\beta$} & \boldsymbol{$df$} & \textbf{SE} & \boldsymbol{$t$} & \boldsymbol{$p$} & \textbf{Predictor} & \boldsymbol{$\beta$} & \boldsymbol{$df$} & \textbf{SE} & \boldsymbol{$t$} & \boldsymbol{$p$}\\
\midrule
(Intercept) & 0.513 & 43.2  & 0.018 & 28.60 & 0.0000 & (Intercept) & 0.467 & 41.1  & 0.019 & 24.67 & 0.0000 \\
shift & -0.035 & 1990.0 & 0.008 & -4.46 & 0.0000 & shift & 0.010 & 1990.0& 0.007 & 1.37 & 0.1722 \\
L & -0.020 & 40.2  & 0.020 & -1.01 & 0.3114 & L & 0.033 & 37.0  & 0.024 & 1.37 & 0.1709 \\
C & 0.030 & 87.5  & 0.011 & 2.74 & 0.0061 & C & -0.000 & 59.6  & 0.013 & -0.01 & 0.9881 \\
a & -0.089 & 44.1  & 0.012 & -7.10 & 0.0000 & a & 0.027 & 44.8 & 0.011 & 2.35 & 0.0188 \\
b & -0.154 & 39.5  & 0.016 & -9.33 & 0.0000 & b & 0.056 & 43.3  & 0.013 & 4.33 & 0.0000 \\

L:C & -0.037 & 1990.0& 0.008 & -4.67 & 0.0000 & L:C & 0.008 & 1990.0 & 0.007 & 1.03 & 0.3016 \\

L:a & 0.011 & 1990.0 & 0.008 & 1.49 & 0.1367 & L:a & -0.007 & 1990.0 & 0.007 & -0.96 & 0.3365 \\
L:b & 0.007 & 1990.0& 0.006 & 1.19 & 0.2337 & L:b & -0.012 & 1990.0 & 0.006 & -2.16 & 0.0310 \\

C:a & 0.024 & 1990.0 & 0.007 & 3.40 & 0.0007 & C:a & -0.001 & 1990.0 & 0.007 & -0.16 & 0.8707 \\
C:b & 0.048 & 1990.0 & 0.006 & 8.63 & 0.0000 & C:b & -0.017 & 1990.0 & 0.005 & -3.13 & 0.0018 \\

L:C:a & 0.005 & 1990.0 & 0.008 & 0.67 & 0.5057 & L:C:a & 0.006 & 1990.0 & 0.008 & 0.83 & 0.4039 \\
L:C:b & -0.000 & 1990.0 & 0.006 & -0.06 & 0.9514 & L:C:b & 0.003 & 1990.0 & 0.006 & 0.51 & 0.6088 \\

L:shift & -0.008 & 1990.0 & 0.008 & -1.00 & 0.3171 & L:shift & 0.033 & 1990.0 & 0.007 & 4.47 & 0.0000 \\
C:shift & 0.012 & 1990.0 & 0.009 & 1.40 & 0.1612 & C:shift & -0.025 &1990.0 & 0.008 & -3.03 & 0.0024 \\

a:shift & 0.005 & 1990.0 & 0.007 & 0.75 & 0.4524 & a:shift & -0.003 & 1990.0 & 0.007 & -0.41 & 0.6788 \\
b:shift & 0.025 & 1990.0 & 0.007 & 3.83 & 0.0001 & b:shift & -0.009 & 1990.0 & 0.006 & -1.42 & 0.1546 \\
L:C:shift & 0.021 & 1990.0 & 0.008 & 2.71 & 0.0068 & L:C:shift & -0.016 & 1990.0 & 0.007 & -2.14 & 0.0321 \\

L:a:shift & 0.007 & 1990.0 & 0.008 & 0.90 & 0.3676 & L:a:shift & -0.001 & 1990.0& 0.007 & -0.11 & 0.9089 \\
L:b:shift & 0.007 & 1990.0 & 0.006 & 1.17 & 0.2421 & L:b:shift & -0.009 & 1990.0 & 0.006 & -1.57 & 0.1154 \\

C:a:shift & 0.005 & 1990.0 & 0.007 & 0.70 & 0.4857 & C:a:shift & 0.006 & 1990.0 & 0.007 & 0.95 & 0.3406 \\

C:b:shift & -0.016 & 1990.0 & 0.006 & -2.93 & 0.0034 & C:b:shift & 0.004 & 1990.0 & 0.005 & 0.84 & 0.4031 \\
L:C:a:shift & -0.008 & 1990.0 & 0.008 & -0.93 & 0.3516 & L:C:a:shift & -0.002 &1990.0 & 0.008 & -0.30 & 0.7627 \\
L:C:b:shift & -0.006 & 1990.0 & 0.006 & -0.96 & 0.3354 & L:C:b:shift & 0.017 & 1990.0 & 0.006 & 3.05 & 0.0023 \\
\hline
\end{tabular}
\label{tab:exp2-lmer-results-2col}
\end{table*}

\begin{table*}[t!] % Or [p] for a "page of floats"
\centering

% --- First Table ---
\caption{Experiment 2 mixed effects model results using colorimetric predictors derived from color scales. Predictors were $z$-scored. (Stars are omitted from L*, C*, a*, and b* for table readability.)}
\label{tab:exp2-scales-lmer-results-2col}
\small
\begin{tabular}{lccccc|lccccc}
\midrule
& & \textbf{Positive} & & & & & & \textbf{Negative} & & \\
\midrule
\textbf{Predictor} & \boldsymbol{$\beta$} & \boldsymbol{$df$} & \textbf{SE} & \boldsymbol{$t$} & \boldsymbol{$p$} & \textbf{Predictor} & \boldsymbol{$\beta$} & \boldsymbol{$df$} & \textbf{SE} & \boldsymbol{$t$} & \boldsymbol{$p$}\\
\midrule
(Intercept) & 0.516 & 55.9  & 0.013 & 39.54 & 0.0000 & (Intercept) & 0.475 & 45.6  & 0.017 & 28.68 & 0.0000 \\
L & 0.102 & 49.4  & 0.014 & 7.15 & 0.0000 & L & -0.102 & 50.4 & 0.014 & -7.53 & 0.0000 \\
C & 0.059 & 129.0  & 0.009 & 6.27 & 0.0000 & C & -0.054 & 99.0  & 0.010 & -5.28 & 0.0000 \\
a & -0.007 & 51.3  & 0.011 & -0.64 & 0.5227 & a & 0.036 & 44.1  & 0.013 & 2.79 & 0.0053 \\
b & -0.064 & 42.9  & 0.014 & -4.57 & 0.0000 & b & 0.080 & 46.5  & 0.012 & 6.56 & 0.0000 \\

L:C & -0.022 & 2000.0 & 0.008 & -2.90 & 0.0037 & L:C & 0.022 & 2000.0 & 0.008 & 2.95 & 0.0032 \\
L:a & -0.004 & 2000.0 & 0.007 & -0.58 & 0.5614 & L:a & -0.009 & 2000.0 & 0.007 & -1.21 & 0.2247 \\
L:b & 0.003 & 2000.0 & 0.006 & 0.50 & 0.6171 & L:b & -0.006 & 2000.0 & 0.006 & -1.01 & 0.3130 \\

C:a & 0.034 & 2000.0 & 0.007 & 5.10 & 0.0000 & C:a & -0.044 & 2000.0 & 0.007 & -6.55 & 0.0000 \\
C:b & 0.032 & 2000.0 & 0.005 & 5.96 & 0.0000 & C:b & -0.039 & 2000.0 & 0.005 & -7.29 & 0.0000 \\
L:C:a & 0.010 & 2000.0 & 0.008 & 1.29 & 0.1957 & L:C:a & -0.006 & 2000.0 & 0.008 & -0.75 & 0.4520 \\
L:C:b & -0.018 & 2000.0 & 0.006 & -3.20 & 0.0014 & L:C:b & 0.011 & 2000.0 & 0.006 & 1.93 & 0.0540 \\
\toprule
& & \textbf{Calm} & & & & & & \textbf{Startling} & & \\
\midrule
\textbf{Predictor} & \boldsymbol{$\beta$} & \boldsymbol{$df$} & \textbf{SE} & \boldsymbol{$t$} & \boldsymbol{$p$} & \textbf{Predictor} & \boldsymbol{$\beta$} & \boldsymbol{$df$} & \textbf{SE} & \boldsymbol{$t$} & \boldsymbol{$p$}\\
\midrule
(Intercept) & 0.513 & 45.6  & 0.018 & 28.91 & 0.0000 & (Intercept) & 0.467 & 42.9  & 0.019 & 24.97 & 0.0000 \\
L & -0.020 & 41.9  & 0.019 & -1.02 & 0.3060 & L & 0.033 & 48.4  & 0.024 & 1.39 & 0.1655 \\
C & 0.030 & 94.3  & 0.011 & 2.74 & 0.0062 & C & -0.000 & 63.2  & 0.013 & -0.01 & 0.9881 \\
a & -0.089 & 46.3  & 0.012 & -7.16 & 0.0000 & a & 0.027 & 47.0  & 0.011 & 2.37 & 0.0177 \\
b & -0.154 & 41.2 & 0.016 & -9.45 & 0.0000 & b & 0.056 & 45.3  & 0.013 & 4.38 & 0.0000 \\

L:C & -0.037 & 2010.0 & 0.008 & -4.59 & 0.0000 & L:C & 0.008 & 2010.0 & 0.008 & 1.02 & 0.3087 \\
L:a & 0.011 & 2010.0 & 0.008 & 1.46 & 0.1435 & L:a & -0.007 & 2010.0 & 0.007 & -0.95 & 0.3435 \\
L:b & 0.007 & 2010.0 & 0.006 & 1.17 & 0.2418 & L:b & -0.012 & 2010.0 & 0.006 & -2.13 & 0.0336 \\
C:a & 0.024 & 2010.0 & 0.007 & 3.34 & 0.0008 & C:a & -0.001 & 2010.0 & 0.007 & -0.16 & 0.8726 \\
C:b & 0.048 & 2010.0 & 0.006 & 8.48 & 0.0000 & C:b & -0.017 & 2010.0 & 0.005 & -3.08 & 0.0021 \\
L:C:a & 0.005 & 2010.0 & 0.008 & 0.65 & 0.5130 & L:C:a & 0.006 & 2010.0 & 0.008 & 0.82 & 0.4108 \\
L:C:b & -0.000 & 2010.0 & 0.006 & -0.06 & 0.9522 & L:C:b & 0.003 & 2010.0 & 0.006 & 0.50 & 0.6140 \\
\hline
\end{tabular}

\vspace{.5cm} % Adjust this value to get the desired spacing between tables

% --- Second Table ---
\caption{Experiment 2 mixed effects model results using colorimetric predictors derived from colormap images. Predictors were $z$-scored. (Stars are omitted from L*, C*, a*, and b* for table readability.)}
\label{tab:exp2-maps-lmer-results-2col}
\begin{tabular}{lccccc|lcccccc}
\midrule
& & \textbf{Positive} & & & & & & \textbf{Negative} & & \\
\midrule
\textbf{Predictor} & \boldsymbol{$\beta$} & \boldsymbol{$df$} & \textbf{SE} & \boldsymbol{$t$} & \boldsymbol{$p$} & \textbf{Predictor} & \boldsymbol{$\beta$} & \boldsymbol{$df$} & \textbf{SE} & \boldsymbol{$t$} & \boldsymbol{$p$}\\
\midrule
(Intercept) & 0.510 & 38.2  & 0.012 & 43.10 & 0.0000 & (Intercept) & 0.477 & 36.2 & 0.016 & 30.51 & 0.0000 \\
L & 0.113 & 38.3  & 0.015 & 7.46 & 0.0000 & L & -0.114 & 37.9  & 0.015 & -7.39 & 0.0000 \\
C & 0.080 & 44.4  & 0.009 & 8.97 & 0.0000 & C & -0.072 & 43.0  & 0.009 & -7.87 & 0.0000 \\
a & -0.017 & 42.8  & 0.010 & -1.65 & 0.0980 & a & 0.044 & 39.5  & 0.013 & 3.56 & 0.0004 \\
b & -0.101 & 52.0 & 0.015 & -6.85 & 0.0000 & b & 0.114 & 60.6  & 0.013 & 9.05 & 0.0000 \\

L:C & -0.015 & 2000.0 & 0.006 & -2.60 & 0.0094 & L:C & 0.016 & 2000.0 & 0.006 & 2.83 & 0.0047 \\
L:a & 0.013 & 2000.0 & 0.006 & 2.15 & 0.0317 & L:a & -0.026 & 2000.0 & 0.006 & -4.46 & 0.0000 \\
L:b & 0.008 & 2000.0& 0.008 & 1.07 & 0.2854 & L:b & -0.014 & 2000.0 & 0.007 & -1.94 & 0.0528 \\

C:a & 0.044 & 2000.0 & 0.006 & 7.21 & 0.0000 & C:a & -0.050 & 2000.0 & 0.006 & -8.37 & 0.0000 \\
C:b & 0.054 & 2000.0 & 0.006 & 8.56 & 0.0000 & C:b & -0.056 & 2000.0 & 0.006 & -9.18 & 0.0000 \\
L:C:a & -0.010 & 2000.0 & 0.010 & -1.03 & 0.3024 & L:C:a & 0.002 & 2000.0 & 0.010 & 0.16 & 0.8734 \\
L:C:b & -0.025 & 2000.0 & 0.006 & -4.07 & 0.0000 & L:C:b & 0.014 & 2000.0 & 0.006 & 2.36 & 0.0184 \\
\toprule
& & \textbf{Calm} & & & & & & \textbf{Startling} & & \\
\midrule
\textbf{Predictor} & \boldsymbol{$\beta$} & \boldsymbol{$df$} & \textbf{SE} & \boldsymbol{$t$} & \boldsymbol{$p$} & \textbf{Predictor} & \boldsymbol{$\beta$} & \boldsymbol{$df$} & \textbf{SE} & \boldsymbol{$t$} & \boldsymbol{$p$}\\
\midrule
(Intercept) & 0.502 & 36.2  & 0.017 & 29.94 & 0.0000 & (Intercept) & 0.463 & 35.8  & 0.018 & 25.93 & 0.0000 \\
L & -0.042 & 36.1  & 0.022 & -1.91 & 0.0563 & L & 0.042 & 35.6  & 0.024 & 1.71 & 0.0876 \\
C & 0.045 & 41.8  & 0.010 & 4.34 & 0.0000 & C & -0.012 & 38.5  & 0.013 & -0.95 & 0.3438 \\
a & -0.103 & 40.5  & 0.012 & -8.45 & 0.0000 & a & 0.034 & 40.9  & 0.012 & 2.90 & 0.0037 \\
b & -0.186 & 47.2  & 0.017 & -10.93 & 0.0000 & b & 0.070 & 56.1  & 0.014 & 4.99 & 0.0000 \\

L:C & -0.032 & 2010.0 & 0.006 & -5.27 & 0.0000 & L:C & 0.008 & 2010.0 & 0.006 & 1.38 & 0.1686 \\

L:a & 0.018 & 2010.0 & 0.006 & 2.85 & 0.0043 & L:a & -0.019 & 2010.0 & 0.006 & -3.17 & 0.0015 \\
L:b & 0.003 & 2010.0 & 0.008 & 0.39 & 0.6972 & L:b & -0.015 & 2010.0 & 0.008 & -1.93 & 0.0535 \\

C:a & 0.031 & 2010.0 & 0.006 & 4.95 & 0.0000 & C:a & 0.001 & 2010.0 & 0.006 & 0.10 & 0.9184 \\

C:b & 0.071 & 2010.0 & 0.006 & 11.02 & 0.0000 & C:b & -0.030 & 2010.0 & 0.006 & -4.63 & 0.0000 \\
L:C:a & -0.007 & 2010.0 & 0.010 & -0.73 & 0.4642 & L:C:a & -0.001 & 2010.0 & 0.010 & -0.07 & 0.9476 \\
L:C:b & -0.000 & 2010.0 & 0.006 & -0.05 & 0.9568 & L:C:b & 0.008 & 2010.0 & 0.006 & 1.28 & 0.2001 \\
\hline
\end{tabular}
\end{table*}

\end{document}